%% file: wileyNJD-Doc.tex
\documentclass[AMA,STIX1COL]{WileyNJD-v2}

\input{commands.tex}

\usepackage{moreverb}

\newcommand\BibTeX{{\rmfamily B\kern-.05em \textsc{i\kern-.025em b}\kern-.08em
T\kern-.1667em\lower.7ex\hbox{E}\kern-.125emX}}

\articletype{Research Article}%

\received{<day> <Month>, <year>}
\revised{<day> <Month>, <year>}
\accepted{<day> <Month>, <year>}

\begin{document}

\title{Fault-Tolerant Individual Pitch Control of Floating Offshore Wind Turbines via Subspace Predictive Repetitive Control}

\author[1]{Yichao Liu*}

\author[1]{Joeri Frederik}

\author[1]{Riccardo M.G. Ferrari}

\author[2]{Ping Wu}

\author[3]{Sunwei Li}

\author[1]{Jan-Willem van Wingerden}

\authormark{Yichao Liu \textsc{et al}}

\address[1]{\orgdiv{Delft Center for Systems and Control}, \orgname{Delft University of Technology}, \orgaddress{\state{Delft}, \country{The Netherlands}}}

\address[2]{\orgdiv{Faculty of Mechanical Engineering \& Automation}, \orgname{Zhejiang Sci-Tech University}, \orgaddress{\state{Hangzhou}, \country{China}}}

\address[3]{\orgdiv{Shenzhen International Graduate School}, \orgname{Tsinghua University}, \orgaddress{\state{Shenzhen}, \country{China}}}

\corres{*Yichao Liu, Corresponding address. \email{yichaoliu629@gmail.com}}

\presentaddress{Delft Center for Systems and Control, Delft University of Technology, Delft, The Netherlands}

\abstract[Abstract]{\input{sections/0_abs}}



\maketitle


\section{Introduction}\label{sec:1}
\input{sections/1_introduction}

\section{Floating offshore wind turbine model} \label{sec:2}
\input{sections/2_description}

\section{SPRC--rationale behind the data-driven adaptive FTIPC}\label{sec:3}
\input{sections/3_model}

\section{Case study}\label{sec:4}
\input{sections/4_simulation}

\section{Conclusions}\label{sec:5}
\input{sections/5_conclusion}

\bibliography{WileyNJD-AMA}%
\input{sections/Tables}

\end{document}

%% file: commands.tex
\usepackage{graphicx,keyval,trig,url} 
\usepackage[T1]{fontenc}     
\usepackage[utf8]{inputenc}	
\usepackage{graphicx}

\usepackage{amsthm}
\usepackage{amssymb}
\usepackage{amsmath}
\usepackage{mathtools}
\usepackage[mathscr]{eucal}
\usepackage{lipsum}
\usepackage{cuted} 
\usepackage{blindtext}
\usepackage{color}
\usepackage{dsfont}

\usepackage{epstopdf}
\usepackage{bm}   

\usepackage{color}
\usepackage{xcolor}

\newtheorem{rem}{Remark}
\usepackage{multirow}
\usepackage{longtable}

\usepackage{mathtools}

%% file: sections/0_abs.tex
Individual Pitch Control (IPC) is an effective and widely-used strategy to mitigate blade loads in wind turbines. 
However, conventional IPC fails to cope with blade and actuator faults, and this situation may lead to an emergency shutdown and increased maintenance costs.
In this paper, a Fault-Tolerant Individual Pitch Control (FTIPC) scheme is developed to accommodate these faults in Floating Offshore Wind Turbines (FOWTs), based on a Subspace Predictive Repetitive Control (SPRC) approach.
To fulfill this goal, an online subspace identification paradigm is implemented to derive a linear approximation of the FOWT system dynamics. Then, a repetitive control law is formulated to attain load mitigation under operational conditions, both in healthy and faulty conditions.
Since the excitation noise used for the online subspace identification may interfere with the nominal power generation of the wind turbine, a novel excitation technique is developed to restrict excitation at specific frequencies.
Results show that significant load reductions are achieved by FTIPC, while effectively accommodating blade and actuator faults and while restricting the energy of the persistently exciting control action. 


%% file: sections/1_introduction.tex
Nowadays, wind is playing an increasingly important role in the international energy mix \cite{Decastro-2019}. For instance, the global installed capacity of wind power over the world reached about 591\,GW in 2018, with a growth of 9 per cent compared to 2017 \cite{GWEC_2019}.
Following the path explored by onshore wind exploitation, the offshore wind energy picks up momentum in the race in transitioning from conventional fossil fuels to renewable energy. 
In comparison, offshore wind turbines are less intrusive from a visual and acoustic point of view, and guarantee much higher and more steady power generation \cite{Decastro-2019}.

With the depletion of coastal resources, geographic limits and environmental constraints, the deployment of the offshore wind turbines tends to move towards deep waters (i.e., a depth of more than 60\,m) to exploit abundant offshore wind resources \cite{Bento_2019}. 
Floating offshore wind turbines (FOWTs) are the only viable solution for harvesting such deep-water wind resources \cite{Liu-2016}. 
However, FOWTs might be subjected to complex and extreme wind and wave loads due to the harsh ocean conditions, which can result in unexpected mechanical and electric faults \cite{Carroll_2016,Liu_2019}.
Particularly, the pitch control system, which is mainly used to optimize power generation and to alleviate blade loads,
is involved in the biggest proportion ($\sim$21.3\,\%) of the overall failure rate for wind turbines \cite{Jiang_2014}. 
Actually, pitch actuators may be occasionally subjected to \emph{severe} faults (e.g., an abrupt Pitch Actuator Stuck (PAS) fault), which leads to a complete loss of control authority, as well as \emph{non-severe} ones, (e.g. Pitch Actuator Degradation (PAD)) \cite{Li_2018}. 
Both these faults may induce unbalanced aerodynamic loads acting on the rotor, which is called aerodynamic asymmetry.
In addition, another cause of aerodynamic asymmetry is the \emph{abrupt structural faults of rotor blades}, which can be induced by the effects of blade cracking, debonding/delamination or fiber breakage \cite{Pawar-2007}.
The issue associated with aerodynamic asymmetry becomes more prominent in FOWTs than land-based counterparts due to the increasing rated power, which implies larger and more complex rotor blades \cite{Faulstich-2011}.

With this in mind, designing both a reliable pitch control and a Fault-Tolerant Control (FTC) strategy is of crucial importance in the development of FOWTs. 
Once a fault occurs, FTC will be activated to assist pitch control strategies in maintaining basic system performances by alleviating the aerodynamic asymmetries and preventing faults from developing into serious failure. 
Currently, a great variety of FTC strategies have been developed in previous studies.
For instance, linear parameter varying \cite{Sloth_2010} and model predictive control based FTC \cite{Yang_2012} have been developed to accommodate pitch actuator faults induced by air contained in the oil of the hydraulic system. 
Furthermore, a control allocation based approach \cite{Kim_2012} {\color{black}was} developed to compensate for the effect of non-severe faulty actuators by implementing a reconfiguration law.
{\color{black}
More recently, adaptive sliding mode observer \cite{Liu_2018, Lan_2018}, unknown input observer \cite{Liu_2020} and model-based FTC scheme \cite{Badihi_2020} were utilized to design the Fault-Tolerant Individual Pitch Control (FTIPC) in the literature.
In addition, other} techniques such as barrier Lyapunov functions \cite{Li_2018}, back propagation neural networks \cite{Rahimilarki_2019}, fuzzy models \cite{badihi2018fault}{\color{black}, sliding mode observer \cite{Rahnavard_2019} were proposed for detecting and accommodating pitch faults in the last few years.}

Nevertheless, the following issues still require further investigations:
\begin{enumerate}
    \item All the FTC strategies developed in previous researches, to the best of authors' knowledge, only address PAD type of faults, without consideration of PAS type of faults.
    In general, the preferred way to address the PAS type faults is through a safe and fast shutdown of the wind turbine \cite{Cho_2018}. 
    However, as PAS type of faults occur frequently \cite{Ribrant_2006}, such a shutdown solution may lead to unnecessarily high Operation and Maintenance (O\&M) costs.
    
    \item Fault accommodation for blade faults are barely considered in the existing FTC strategies, even though aerodynamic asymmetry induced by blade faults has received extensive concern and investigated in the literature \cite{Gong_2010}.
    The reason for this is that there is currently no wind turbine model including the fault scenarios of the rotor blades, nor are these implementable in widely-used simulation packages such as NREL's Fatigue, Aerodynamics, Structures, and Turbulence (FAST) model \cite{Jonkman-2005}.
    
    \item Compared to the conventional baseline classical Collective Pitch Control (CPC) and Individual Pitch Control (IPC) strategies, the existing FTC strategies require much higher activities of pitch actuation, which will cause significant cyclic fatigue loads on the pitch actuators and lead to premature failure of these turbine components.
    
    \item It is usually assumed that fatigue damage is mainly caused by One-Per (1P) deterministic loads, and thereby most of publications only apply 1P revolution cyclic load mitigation. However, higher harmonics, e.g. 2P, 3P, etc. may have significant effect on large-scale wind turbines \cite{Jackson_2016}.
    
    \item Most importantly, only land-based wind turbines were considered in the design of FTC strategies \cite{odgaard2013wind}, thus excluding the aerodynamics, hydrodynamics, actuation, mooring and structural dynamic peculiarities of FOWTs. Therefore, such methods may be underperforming or not reliable when applied directly to FOWTs.
\end{enumerate}

To approach the dearth of the FTC strategies against the blade and pitch actuator faults, a data-driven adaptive {\color{black}FTIPC} is developed for FOWTs in this paper.
More specifically, it is based on a fully data-driven method called \emph{Subspace Predictive Repetitive Control (SPRC) with restricted excitation}. 
The concept of SPRC was initially proposed by van Wingerden and Hulskamp \cite{Wingerden_2011}.
{\color{black}
It showed promising results in simulations \cite{Navalkar_2014} and in wind tunnel experiments \cite{Navalkar_2015, frederik2018}.}
It consists of online subspace identification and Repetitive Control (RC).
The subspace identification \cite{vanderVeen_2013}, is used to obtain a linear approximation of wind {\color{black}turbines} online. Based on this model, a predictive RC law can be formulated to minimize the blade loads under operational conditions.

With the goal of developing the data-driven adaptive FTIPC, this paper contributes in the following two aspects.
First, the SPRC approach is tailored to make the control law {\color{black}adapted} to the faulty conditions.
This is achieved by automatically identifying the Markov matrix for each blade.
{\color{black}Therefore, the fault dynamics can be included in the linear model once a blade or pitch actuator suffers from faults.}
An adaptive fault-tolerant RC law is then formulated to mitigate the fault-induced loads.

Another key contribution of this paper is the restricted excitation in SPRC.
For the purpose of the subspace identification, the FOWT dynamics are, in general, excited persistently by {\color{black}the} external random noise \cite{vanderVeen_2013}. 
However, such a persistent excitation may affect the nominal power production of the operational wind turbine. Even worse, it will induce unexpected degradation of the system performance and operation interruption if a fault occurs. 
To deal with this problem, currently the common way is to switch the subspace identification offline after the model parameters are successfully estimated \cite{Wingerden_2011}.
However, such an 'offline-switch' method is infeasible for FTC once a fault occurs and the system dynamics are changed with it.
Therefore, a novel restricted excitation technique is developed in this paper to guarantee that the energy of the persistently exciting control input only concentrates at frequencies of interests, i.e. 1P and 2P frequencies.

The effectiveness of the proposed FTIPC will be illustrated via a series of case studies involving a 10MW FOWT model developed by Technical University of Denmark (DTU)~\cite{Bak_2013} and Stuttgart Wind Energy (SWE) institute~\cite{Lemmer_2016}.

The remainder of the paper is organized as follows. Section \ref{sec:2} presents the 10MW FOWT model and the simulation environment. In Section \ref{sec:3}, the theoretical framework of FTIPC is detailed. Next, a series of case studies utilizing the high-fidelity simulation package is implemented in Section \ref{sec:4}. Section \ref{sec:5} draws the conclusions.

%% file: sections/2_description.tex
\begin{figure}
\centering 
\includegraphics[width=0.4\columnwidth]{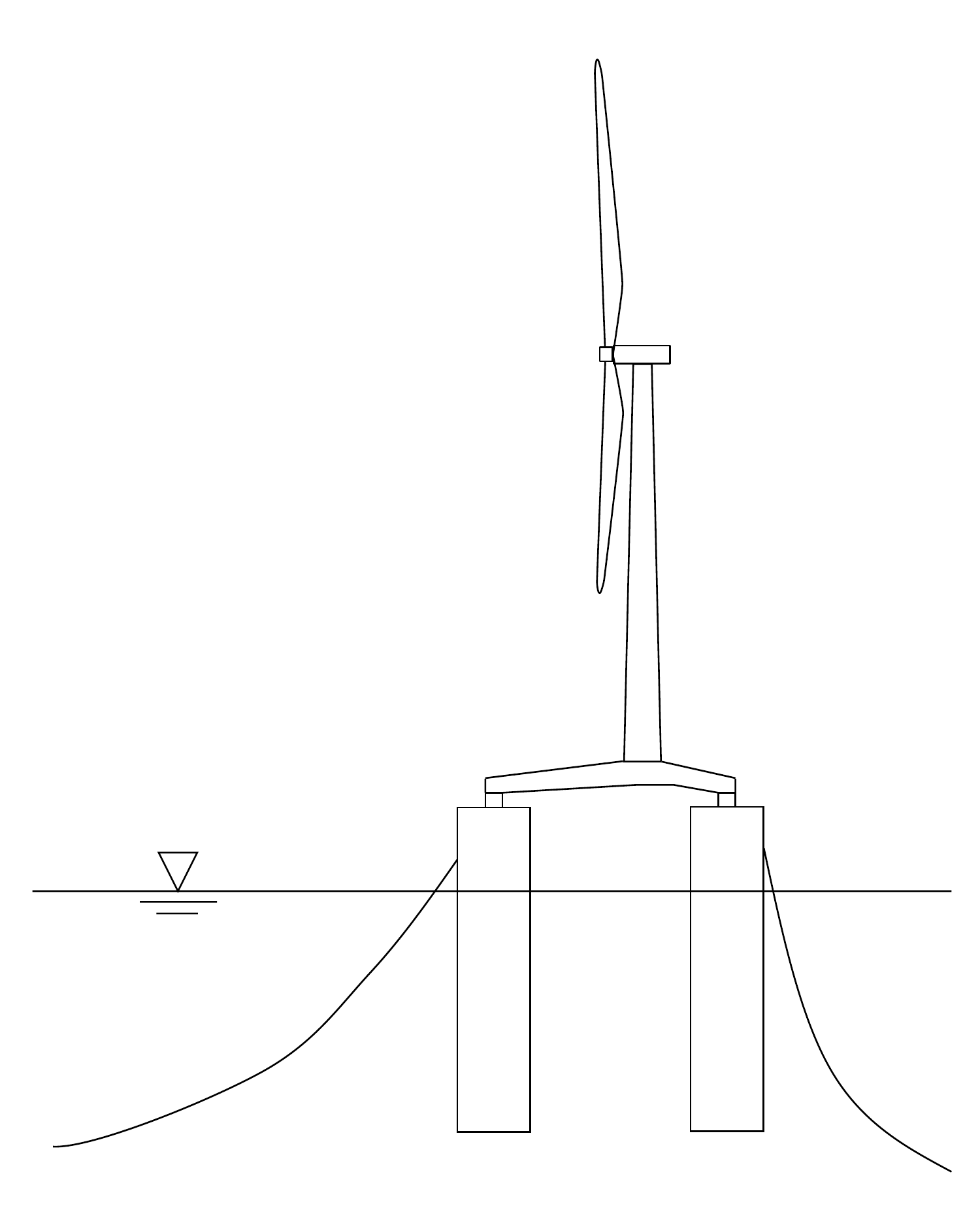}
\caption{Lateral view of the DTU 10MW wind turbine with the TripleSpar floating platform.}
\label{Pic_drawing} %
\end{figure}

The FOWT model used for the case studies is introduced in this section. 
The model is based on the DTU 10MW three-bladed variable speed reference wind turbine combined with the TripleSpar floating platform, as shown in Figure~\ref{Pic_drawing} and Table~\ref{table:FOWT}.

\begin{figure}
\centering 
\includegraphics[width=0.6\columnwidth]{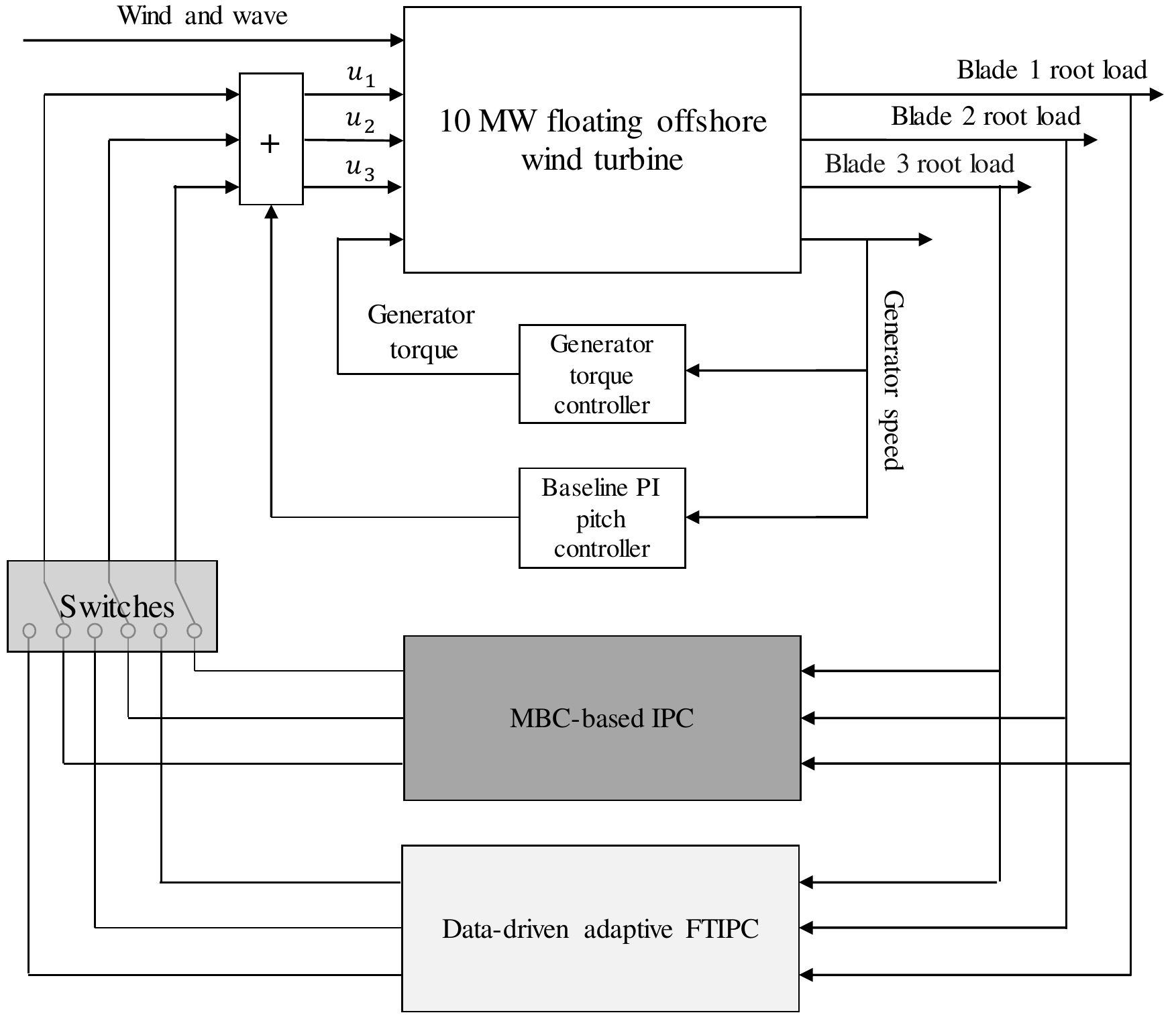}
\caption{Block scheme of the controller and loop of the 10MW FOWT model. The aero-hydro-structural dynamic part is simulated in FAST code while the turbine control part is implemented in Simulink. Three pitch control strategies, i.e. baseline CPC, MBC-based IPC and data-driven adaptive FTIPC are included for comparisons.}
\label{Pic_block} %
\end{figure}

The implementation of the case studies focusing on the 10MW FOWT model is portrayed in Figure~\ref{Pic_block}. It consists of an aero-hydro-structural dynamic part simulated in the widely-used FAST numerical package \cite{Jonkman-2005}, while the wind turbine control part is implemented in \emph{Simulink}.
In particular, the pitch control strategies are divided into{\color{black}:} 1) baseline control using CPC \cite{Jonkman-2005}, which encompasses a white block in Figure~\ref{Pic_block}, 2) Multi-Blade Coordinate (MBC)-based IPC \cite{Mulders_2019}, which is represented by a dark grey block, 3) Data-driven adaptive FTIPC developed in this paper, which is characterized by a light grey block and will be elaborated in Section \ref{sec:3}.
Note that CPC is based on the classical gain-scheduled Proportional Integral (PI) control \cite{Boukhezzar_2007} and regulates the pitch angles of all blades synchronously.
In MBC-based IPC, instead, the pitch angle of each blade is regulated independently with the aid of the classical Coleman transformation \cite{Mulders_2019}. 

Regarding the aero-hydro-structural dynamic part, the 10MW FOWT can be modelled by the following discrete-time system,
\begin{align}
\begin{cases}
 x_{k+1} \!\!\! &= A^0x_k+\rho(x_k,u_k,\phi^x(u_k,\vartheta^x)) +\eta^{x}(x_k,u_k,k)  \\
 y_k \!\!\! &= C^0x_k +\eta^{y}(x_k,u_k,k)
\end{cases} \, ,
\label{eq:FOWT_DYNAMICS}
\end{align}
where $x_k\in\mathbb{R}^n$, $u_k\in\mathbb{R}^r$, $y_k\in\mathbb{R}^l$ represent the state, the control input and the measurement output vectors at the time instant $k$, respectively.
The content of $u$ {\color{black}represents} the vector of the three blade pitch inputs {\color{black}and $r$ is the number of the inputs ($u^{(r)}$, $r=3$).}
{\color{black}The content of $y$ denotes the three blade load signals while $l$ is the number of the outputs ($y^{(l)}$, $l=3$)}, e.g. Out-of-Plane bending moment (MOoP) at the blade root measured by strain gauges or equivalent sensors.
The system matrix $A^0\in\mathbb{R}^{n\times{n}}$ and the vector field $\rho:\mathbb{R}^n\times\mathbb{R}^r\mapsto\mathbb{R}^n$ denote the nominal linear and nonlinear parts of the 10MW FOWT healthy dynamics while $C^0\in\mathbb{R}^{l\times{n}}$ is the nominal output matrix. The unavoidable modelling uncertainties and output disturbances are characterized by the unknown but bounded function $\eta^x:\mathbb{R}^n\times\mathbb{R}^r\times\mathbb{R}\mapsto\mathbb{R}^n$ and $\eta^y:\mathbb{R}^n\times\mathbb{R}^r\times\mathbb{R}\mapsto\mathbb{R}^l$.

Furthermore, the term $\phi^x(u_k,\vartheta^x)$ describes the changes in the state equation \eqref{eq:FOWT_DYNAMICS}, due to the occurrence of the blade and pitch actuator faults at the faulty time index $k_0$. Particularly, the PAS type of faults can be modelled by the following equation,
\begin{equation}
\begin{gathered}
\phi^{x}\,\,\,=-u_{(f)}+\vartheta^{x}_{(\text{PAS})}e_{(f)} 
\, ,
\end{gathered}
\label{eq:PAS}
\end{equation}
{\color{black}where $\vartheta^{x}_{(\text{PAS})}$ equals the value of pitch angle of the $f^{\text{th}}$ blade caused by the stuck actuator.
In this case, once the pitch actuator is stuck, it can not be moved anymore. 
The pitch control input from the faulty actuator is fixed and set as $\vartheta^{x}_{(\text{PAS})}$.
Therefore, the faulty pitch actuator does not work and such a severe PAS type of fault can not be compensated.
The objective of the developed FTIPC is to alleviate the negative effects of the fault by formulating the pitch control inputs for the remaining healthy blades.
The PAD type of faults are represented as,}
\begin{equation}
\begin{gathered}
{\color{black}\phi^{x}\,\,\,=-\vartheta^x_{(\text{PAD})}u_{(f)}e_{(f)} }
\, ,
\end{gathered}
\label{eq:PAD}
\end{equation}
{\color{black}where $\vartheta^{x}_{(\text{PAD})}$ is a scale factor: $0<\vartheta^{x}_{(\text{PAD})}\leq 1$.}
In addition, $e_{(f)}$ represents an all zero-column vector of suitable size having a single 1 in its $f^{\text{th}}$ position ($f=1,2,3$). 
{\color{black}In this case, one of the pitch actuators suffers from a scaling type of fault.
Since it is a non-severe PAD type of fault, all the pitch actuators are still working and the fault can be compensated by the developed FTIPC approach.}

With respect to the structural faults of the rotor blade, it can be modelled as,
\begin{equation}
\begin{gathered}
\phi^{x}= \Delta \rho(x,u,\vartheta^{x}_{\text{B}}) 
\, ,
\end{gathered}
\label{eq:BF}
\end{equation}
where $\phi^{x}$ describes structural faults of the rotor blade that affect the nonlinear dynamic function $\rho$. $\Delta \rho(x,u,\vartheta^{x}_{\text{B}})$ denotes the change in the nonlinear part of the dynamics from its nominal behaviour to a faulty behaviour characterized by the parameters vector $\vartheta^{x}_{\text{B}}$. 
The vector includes specific values of structural parameters which are caused by the structural faults, e.g., reduced blade stiffness. 
The fault function $\vartheta^{x}_{\text{B}} = a\cdot \vartheta^{x}$ is utilized to describe blade faults caused by cracking, debonding/delamination or fiber breakage \cite{Pawar-2007}, where $0 < a\leq 1$ is a scale factor determining the reduction of the blade stiffness.
{\color{black}Similarly, in this case all the pitch actuators are still working and the fault can be compensated by the developed FTIPC approach.}
In order to properly produce the abrupt rotor blade faults in the 10MW FOWT model, the corresponding faulty mode shapes of the blades, which are dependent on $\vartheta^{x}_{\text{B}}$, are estimated using the tool Modes\footnote{Modes: a simple mode-shape generator for towers and rotating blades. https://nwtc.nrel.gov/Modes.} and then fed into the turbine FAST model. 

{\color{black}It should be noted that only one blade/actuator is subjected to the specific fault at each fault scenario in the present study.}

%% file: sections/3_model.tex
In this section the SPRC theory, which is used to develop the data-driven adaptive FTIPC, is introduced. 
First, a discrete-time Linear Time Invariant (LTI) system along with an output predictor is formulated, on which the RC law will be later formulated. 
Then, the parameters of the linear approximation are identified online via subspace identification, during which the FOWT dynamics are persistently excited only at the frequencies of interest, i.e. 1P and 2P frequencies, via the restricted excitation technique developed in this paper.
This approach will be described in detail in Section 3.3.
Based on this, the estimated model parameters are utilized to derive the RC law that minimizes the blade loads in operational conditions.
{\color{black}Once a blade or pitch actuator is subjected to faults, this control strategy will include the fault dynamics into the linear model automatically by identifying the model parameters for each blade.} 
Finally, the adaptive fault-tolerant RC inputs are formulated to mitigate the fault-induced loads.
\subsection{Online subspace identification}
{\color{black}The online subspace identification in SPRC is based on the idea originally from a so-called Predictor Based Subspace IDentification (PBSID) approach \cite{Chiuso_2007}.}
In {\color{black}detail}, the 10MW FOWT system dynamics governed by equation~\eqref{eq:FOWT_DYNAMICS} can be approximated by a linear model affected by unknown periodic disturbances {\color{black}and faults \cite{Houtzager_2013} in innovation form} as,
\begin{equation}
\begin{cases}
\label{eq:LTI}
x_{k+1} \!\! &= Ax_k+B(u_k+\phi^x)+Ed_k+Le_k \\
y_k \!\! &= Cx_k+Fd_k+e_k
\end{cases} \, ,
\end{equation}
where $x_k\in\mathbb{R}^n$, $u_k\in\mathbb{R}^r$ and $y_k\in\mathbb{R}^l$ represent the state, control input and output vectors.
In the present case, $r = l = 3$, $u_k$ and $y_k$ indicate the blades pitch angles and the blade loads, i.e. MOoP, at discrete time index $k$, respectively. 
Furthermore, $d_k\in\mathbb{R}^m$ represents the periodic disturbance component of the loads at the blade root, {\color{black}which is caused by the tower shadow, wind shear, yawed error and gravity \cite{Houtzager_2013}.}
$e_k\in\mathbb{R}^l$ is the zero-mean white innovation process or the aperiodic component of the blade loads.
Other matrices $A\in\mathbb{R}^{n\times n}$, $B\in\mathbb{R}^{n\times r}$, $C\in\mathbb{R}^{l\times n}$, $L\in\mathbb{R}^{n\times l}$, $E\in\mathbb{R}^{n\times m}$ and $F\in\mathbb{R}^{l\times m}$ are the state transition, input, output, observer, periodic noise input and periodic noise direct feed-through matrices, respectively. Under the healthy condition ($0\leq{k} < k_0$), $\phi^x=0$ holds.
\begin{rem}
{\color{black}
Equation~\eqref{eq:LTI} is the so-called innovation form of the state space model which is compatible with observations and a Kalman filter prediction \cite{Kunsch_2000}.
With this formulation, the dynamics of the states incorporate the information collected from the measurements.
All components of the state vector whose information can not be obtained from the observations are automatically removed, thus leading to a minimal model of the wind turbine.}
\end{rem}
In predictor form, equation \eqref{eq:LTI} can be reformulated as,
\begin{equation}
\begin{cases}
\label{eq:predictor:f}
x_{k+1} \!\! &= \tilde{A}x_k+B(u_k+\phi^{x})+\tilde{E}d_k+Ly_k \\
y_k \!\! &= Cx_k+Fd_k+e_k
\end{cases} \, ,
\end{equation}
where $\tilde{A}\triangleq{A-LC}$ and $\tilde{E}\triangleq{E-LF}$.
By defining a periodic difference operator $\delta$ as
\begin{align*}
 \delta{d}_k &=d_k-d_{k-P}=0, \\
 \delta{u}_k &= (u_k+\phi^x_k) - (u_{k-P}+\phi^x_{k-P}), \\
 \delta{y}_k &=y_k-y_{k-P},
\end{align*} 
the effect of periodic blade loads $d_k$ on the input-output system can be eliminated. $P$ is the {\color{black}prediction horizon covering the} period of the disturbance, which {\color{black}equals} the blade rotation period.
Based on the definition of $\delta$, equation \eqref{eq:predictor:f} is rewritten as,
\begin{equation}
\begin{cases}
\delta{x}_{k+1} \!\! &= \tilde{A}\delta{x}_k+B\delta{u}_k+L\delta{y}_k \\
\delta{y}_k \!\! &= C\delta{x}_k+\delta{e}_k \, .
\end{cases} 
\label{eq:predictor2} \, .
\end{equation}
Given the past time window with the length of $p$, the following stacked vector $\delta{U}^{(p)}_{k}$ can be defined:
\begin{equation}
\delta{U}^{(p)}_{k}=
\left[ \begin{array}{c}
 u_k-u_{k-P}\\
u_{k+1}-u_{k-P+1} \\
\vdots \\
u_{k+p-1}-u_{k+p-P-1}
\end{array} 
\right ]\, .
\label{eq:stacked u}
\end{equation}
Similarly, the vector $\delta{Y}^{(p)}_{k}$ can be defined. The future state vector $\delta{x}_{k+p}$ can then be introduced based on $\delta{U}^{(p)}_{k}$ and $\delta{Y}^{(p)}_{k}$ as
\begin{equation}
\delta{x}_{k+p}= \tilde{A}^{p}\delta{x}_k+
\left[ \begin{array}{cc}
K^{(p)}_u & K^{(p)}_y 
\end{array} 
\right ]
\left[ \begin{array}{c}
\delta{U}^{(p)}_{k} \\
\delta{Y}^{(p)}_{k} \\
\end{array} 
\right ]\, ,
\label{eq:lifted}
\end{equation}
where $K^{(p)}_u$ and $K^{(p)}_y$ are defined as,
\begin{align*}
&\ K^{(p)}_u=
\left[ \begin{array}{cccc}
\tilde{A}^{p-1}B & \tilde{A}^{p-2}B & \cdots & B 
\end{array} 
\right ]\,,\\ 
&\ K^{(p)}_y=
\left[ \begin{array}{cccc}
\tilde{A}^{p-1}L & \tilde{A}^{p-2}L & \cdots & L
\end{array} 
\right ]\, .
\label{eq:K}
\end{align*}

It is noted that $p$ should be chosen large enough, such that $\tilde{A}^{j}\approx0$ $\forall{j}\geq{p}${\color{black}\cite{Chiuso_2007}}. In such a case, $\delta{x}_{k+p}$ is approximated by
\begin{equation}
\delta{x}_{k+p}=
\left[ \begin{array}{cc}
K^{(p)}_u & K^{(p)}_y 
\end{array} 
\right ]
\left[ \begin{array}{c}
\delta{U}^{(p)}_{k} \\
\delta{Y}^{(p)}_{k} \\
\end{array} 
\right ] \, .
\label{eq:lifted2}
\end{equation}
By combining equations \eqref{eq:predictor2} and \eqref{eq:lifted2}, the approximation of $\delta{y}_{k+p}$ is obtained as
\begin{equation}
\delta{y}_{k+p}=
\underbrace{
\left[ \begin{array}{cc}
CK^{(p)}_u & CK^{(p)}_y 
\end{array} 
\right ]}_{\Xi}
\left[ \begin{array}{c}
\delta{U}^{(p)}_{k} \\
\delta{Y}^{(p)}_{k} \\
\end{array} 
\right ]
+\delta{e_{k+p}} \, .
\label{eq:lifted3}
\end{equation}
It is clear from equation (\ref{eq:lifted3}) that the matrix of coefficients $\left[\begin{array}{cc} CK^{(p)}_u & CK^{(p)}_y\end{array}\right]$, which is defined as Markov matrix $\Xi$, includes all the necessary information on the behaviour of the FOWT and can be estimated based on the input vector $u^{(r)}$ and output vector $y^{(l)}$. 
%
{\color{black}
In order to make FTIPC adapted to different conditions, i.e. healthy and faulty conditions, the assumption is made that the load of blade $n$ is independent from  the blade $m$, where $n$ and $m$ are blade numbers and $n\neq{m}$.
Therefore, the system identification is implemented by obtaining} an online solution of the following Recursive Least-Squares (RLS) optimization problem \cite{vanderVeen_2013}
\begin{equation}
{\color{black}
\hat{\Xi}_{k,(i)}=\text{arg}\min_{\hat{\Xi}_{k,(i)}}\sum_{k=0}^{\infty}\left \| \delta{y}_{k,(i)}-\lambda\hat{\Xi}_{k,(i)}
\left[ \begin{array}{c}
\delta{U}^{(p)}_{k,(i)} \\
\delta{Y}^{(p)}_{k,(i)} \\
\end{array} 
\right ]
\right \| ^2_2 \, ,
}
\label{eq:Markov parameters3}
\end{equation}
where $\lambda$ is a forgetting factor ($0\ll\lambda\leq{1}$) to attenuate the effect of past data, and adapt to the updated system dynamics online. In this paper, a large value, i.e. $\lambda=0.99999$, was selected to guarantee the robustness of the optimization process. 
{\color{black}
$i=1,2,3$ corresponds to the blade number, while $\hat{\Xi}_{k,(i)}$ is the estimate of independent Markov matrix for each blade. Consequently, the optimization process in equation (\ref{eq:Markov parameters3}) is conducted three times at each time instant $k$. Next, $\hat{\Xi}_{k}$ is reformulated as $\hat{\Xi}_{k}=[\hat{\Xi}_{k,(1)}, \hat{\Xi}_{k,(2)}, \hat{\Xi}_{k,(3)}]^T$.
Based on the input and output vectors from the excited 10MW FOWT, the above mentioned RLS optimization is implemented to obtain $\hat{\Xi}_{k}$ via a QR algorithm \cite{Sayed_1998}. Next, the estimates of $\hat{\Xi}_k$ are used to formulate a RC law, as explained in the following subsection.
}
Consequently, $\hat{\Xi}_k$ contains estimates of the following matrices based on RLS optimization, as
\begin{equation}
\hat{\Xi}_k=
\biggl[ 
\begin{array}{cccccccc}
\widehat{C\tilde{A}^{p-1}B} & \widehat{C\tilde{A}^{p-2}B} & \cdots & \widehat{CB} &
\widehat{C\tilde{A}^{p-1}K} & \widehat{C\tilde{A}^{p-2}K} & \cdots & \widehat{CK}
\end{array}
\biggr] \, .
\label{eq:estimates of Markov parameter}
\end{equation}

\subsection{Infinite horizon repetitive control}
The RC law is predicted over $P$, where $P\geq{p}$ and usually $P$ is much larger than $p$. Hence, the output equation is lifted over the period $P$ as
\begin{equation}
\delta{Y}^{(P)}_{k+p}=\tilde{\Gamma}^{(P)}\delta{x}_{k+p}+
\left[ \begin{array}{cc}
\tilde{H}^{(P)} & \tilde{G}^{(P)} \\
\end{array} 
\right ]
\left[ \begin{array}{c}
\delta{U}^{(P)}_{k+p} \\
\delta{Y}^{(P)}_{k+p} \\
\end{array} 
\right ]
\, ,
\label{eq:lift:y}
\end{equation}
where the Toeplitz matrix $\tilde{H}^{(P)}$ is defined as,
\begin{equation}
\tilde{H}^{(P)}=
\left[ \begin{array}{cccc}
0 & 0 & 0 & \cdots  \\
CB & 0 & 0 & \cdots  \\
C\tilde{A}B & CB & 0 & \cdots \\
\vdots & \vdots & \ddots & \vdots \\
C\tilde{A}^{p-1}B & C\tilde{A}^{p-2}B & C\tilde{A}^{p-3}B & \cdots \\
0 & C\tilde{A}^{p-1}B & C\tilde{A}^{p-2}B & \cdots \\
0 & 0 & C\tilde{A}^{p-1}B & \cdots \\
\vdots & \vdots & \ddots & \ddots \\
\end{array} 
\right ]
\, .
\label{eq:H}
\end{equation}
Similarly, $\tilde{G}^{(P)}$ is obtained by replacing $B$ by $L$. The extended observability matrix $\tilde{\Gamma}^{(P)}$ is defined as
\begin{equation}
\tilde{\Gamma}^{(P)}=
\left[ \begin{array}{c}
C \\
C\tilde{A} \\
C\tilde{A}^2 \\
\vdots \\
C\tilde{A}^p \\
0 \\
\vdots \\
0
\end{array} 
\right ]
\, .
\label{eq:Gamma}
\end{equation}
Since this {\color{black}is} a predictor, the white noise sequence $\delta e$ is omitted.
Combining equation (\ref{eq:lift:y}) with (\ref{eq:lifted2}), this can be rewritten as
\begin{equation}
\delta{Y}^{(P)}_{k+P}=\tilde{\Gamma}^{(P)}
\left[ \begin{array}{cc}
K^{(P)}_u & K^{(P)}_y \\
\end{array} 
\right ]
\left[ \begin{array}{c}
\delta{U}^{(P)}_k \\
\delta{Y}^{(P)}_k 
\end{array} 
\right ]
+ 
\left[ \begin{array}{cc}
\tilde{H}^{(P)} & \tilde{G}^{(P)} \\
\end{array} 
\right ]
\left[ \begin{array}{c}
\delta{U}^{(P)}_{k+P} \\
\delta{Y}^{(P)}_{k+P} 
\end{array} 
\right ]
\, .
\label{eq:lift:y2}
\end{equation}
Note that the first $(P-p)\cdot{r}$ columns of $K^{(P)}_u$ and $K^{(P)}_y$ are 0.
According to equation (\ref{eq:lift:y2}), the future output $\delta Y^{(P)}_{k+P}$ is then predicted by $\delta Y^{(P)}_{k}$ and $\delta U^{(P)}_{k}$ over the last period, and input $\delta U^{(P)}_{k+P}$ over the current period as
\begin{equation}
\delta{Y}^{(P)}_{k+P} =
\left[ \begin{array}{cc}
\widehat{\Gamma^{(P)}K^{(P)}_u} & \widehat{\Gamma^{(P)}K^{(P)}_y} \\
\end{array} 
\right ]
\left[ \begin{array}{c}
\delta{U}^{(P)}_k \\
\delta{Y}^{(P)}_k 
\end{array} 
\right ]
+\hat{H}^{(P)}\delta{U}^{(P)}_{(k+P)} \, ,
\label{eq:lift:y2-2}
\end{equation}
where $\Gamma^{(P)}$ and $\hat{H}^{(P)}$ are defined in following equalities:
\begin{align*}
 (I-\tilde{G}^{(P)})^{-1}\tilde{\Gamma}^{(P)} &= \Gamma^{(P)} \\
 (I-\tilde{G}^{(P)})^{-1}\tilde{H}^{(P)} &= \hat{H}^{(P)} 
\end{align*}   \, .

Then the absolute output is to be penalised in the optimization problem, thus making the equation~\eqref{eq:lift:y2-2} expanded as
\begin{equation}
{Y}^{(P)}_{k+P}=
\left[ \begin{array}{ccc}
I_{l\cdot{P}} & \widehat{\Gamma^{(P)}K^{(P)}_u} & \widehat{\Gamma^{(P)}K^{(P)}_y} \\
\end{array} 
\right ]
\left[ \begin{array}{c}
Y^{(P)}_k \\
\delta{U}^{(P)}_k \\
\delta{Y}^{(P)}_k 
\end{array} 
\right ]
+\hat{H}^{(P)}\delta{U}^{(P)}_{(k+P)} \, .
\label{eq:lift:y3}
\end{equation}
Equation \eqref{eq:lift:y3} is reformulated into a state space representation, so that the classical optimal state feedback matrix can be synthesised in a {\color{black}Linear Quadratic Regulator (LQR)} sense \cite{Hallouzi_2006} as
\begin{equation}
\underbrace{
\left[ \begin{array}{c}
Y^{(P)}_{k+P}\\
\delta{U}^{(P)}_{k+P} \\
\delta{Y}^{(P)}_{k+P} \\
\end{array} 
\right ]}_{\hat{K}_{k+P}}
=
\underbrace{
\left[ \begin{array}{ccc}
I_{l\cdot{P}} & \widehat{\Gamma^{(P)}K^{(P)}_u} & \widehat{\Gamma^{(P)}K^{(P)}_y} \\
0_{(r\cdot{P})\times{(l\cdot{P})}} & 0_{r\cdot{P}} & 0_{(r\cdot{P})\times{(l\cdot{P}})} \\
0_{l\cdot{P}} & \widehat{\Gamma^{(P)}K^{(P)}_u} & \widehat{\Gamma^{(P)}K^{(P)}_y} 
\end{array} 
\right]}_{\hat{A}_K}
\underbrace{
\left[ \begin{array}{c}
Y^{(P)}_k \\
\delta{U}^{(P)}_k \\
\delta{Y}^{(P)}_k
\end{array} 
\right]}_{\hat{K}_K}
+
\underbrace{
\left[ \begin{array}{c}
\hat{H}^{(P)} \\
I_{r\cdot{P}} \\
\hat{H}^{(P)}
\end{array} 
\right]}_{\hat{B}_K}
\delta{U}^{(P)}_{k+P}
\, .
\label{eq:state-space form}
\end{equation}
The state transition and input matrices are updated at each discrete time instance $k$. Based on the state space representation, the state feedback controller is implemented according to the Discrete Algebraic Riccati Equation (DARE). 
Then, a basis function projection \cite{Wingerden_2011} is utilized to restrain the spectral content of $U_k$ within the desired frequency range. In addition, such a projection reduces the dimension of DARE that must be solved, thus reducing the computational cost.
In this paper, the blade loads at 1P and 2P frequencies, which are mainly caused by the wind shear, wind turbulence, changes in the inflow wind speed and tower shadow, are considered based on the following transformation matrix
\begin{equation}
\phi=
\underbrace{
\left[ \begin{array}{cccc}
\sin(2\pi/P) & \cos(2\pi/P) & \sin(4\pi/P) & \cos(4\pi/P)\\
\sin(4\pi/P) & \cos(4\pi/P) & \sin(8\pi/P) & \cos(8\pi/P)\\
\vdots & \vdots & \vdots & \vdots\\
\sin(2\pi) & \cos(2\pi) & \sin(4\pi) & \cos(4\pi)
\end{array} 
\right ]}_{U_f} 
\otimes{I}_r
\, ,
\label{eq:with basis function}
\end{equation}
where $U_f$ is defined as the basis function while the symbol $\otimes$ is the Kronecker product.
\begin{rem}\label{rem:basis function}
{\color{black}
One issue should be handled in this approach is the potential variation of rotor speed, due to the occurrence of faults, the wind turbulence or the changes in the inflow wind speed. 
This will lead to a phase shift between input and output.
To deal with this issue}, the rotor azimuth $\psi$, equal to $2\pi{k}/P$ at time instant $k$, is used {\color{black}to reformulate $U_f$} to include rotor speed variations.
{\color{black}
Consequently, equation~\eqref{eq:with basis function} is rewritten as,}
\begin{equation}
{\color{black}
\phi=
\underbrace{
\left[ \begin{array}{cccc}
\sin(\psi) & \cos(\psi) & \sin(2 \psi) & \cos(2 \psi)\\
\end{array} 
\right ]}_{U_f} 
\otimes{I}_r
}
\, ,
\label{eq:with basis function2}
\end{equation}
\end{rem}
By taking a linear combination of the sinusoids of the transformation matrix, the control inputs for the {\color{black}specific frequencies $U^{(P)}_k$} are formulated as
\begin{equation}
U^{(P)}_k=\phi \cdot \theta_j
\, ,
\label{eq:control input}
\end{equation}
where $j=0,1,2,\cdots$ is the rotation count. 
{\color{black}
$\theta_j$ is the transformed coefficients of the pitch control signals $U^{(P)}_k$ which is projected by $\phi$.
The vector ${\color{black}\theta_j} \in\mathbb{R}^{4r}$, determining the amplitudes and phase of the sinusoids, is updated based on equation~\eqref{eq:state-space form} at each $P$.
}
\begin{rem}\label{rem:computational complexicity and time deday}
{\color{black}
It should be stressed that in this approach, the computational complexity is relatively low. The first reason is the basis function projection is effective at reducing the dimension of the linear model. Secondly, the repetitive control law is only computed once per rotation period, which reduces the computational burden. 
Finally, the LQR algorithm used in the repetitive control step is a simple but effective approach, and
 will not induce much computational complexity. 
In terms of the computational time delay, it has been already taken into account in the SPRC algorithm, since the linear model identified by the online subspace identification is based on the discrete-time input and output data.}
\end{rem}
\subsection{Restricted excitation}
In order to obtain a unique solution of the RLS optimization with the adaption to different operation points, e.g. variations in winds and waves or considerations of healthy and faulty conditions, the 10MW FOWT should be persistently excited by external noises online. However, the challenge is that the nominal power production would be significantly affected by such a persistently exciting signal, which is nevertheless necessary for online identification.

Therefore, a restricted excitation technique is developed in this paper that stimulates the system strictly at the frequencies of interest, i.e. 1P, 2P.
In details, the exciting signal is superimposed on top of $\theta_j$ as
\begin{equation}
U^{(P)}_k=\phi(\theta_j+\eta_j)
\, ,
\label{eq:control input2}
\end{equation}
{\color{black}where the vector $\eta_j \in \mathbb{R}^{4r}$ is the filtered pseudo-random binary noise}, which is used to excite system at 1P and 2P frequencies. 
Thanks to the transformation matrix $\phi$ in equation~\eqref{eq:control input2}, the energy of the persistently exciting control input $U^{(P)}_k$ is projected onto the specified 1P and 2P frequencies.

\begin{rem}\label{rem:exciting noise}
In order to guarantee the successful excitation, $\eta$ should be generated in an uncorrelated way with different random seeds for each component of the vector $\theta$. 
\end{rem}
Then, the output is projected onto the subspace that {\color{black}is} defined by the basis vectors as well, as
\begin{equation}
\bar{Y}_k=\phi^{+}Y^{(P)}_k
\, ,
\label{eq:control output}
\end{equation}
where the symbol $+$ denotes the Moore-Penrose pseudo-inverse. Based on the matrix transformation, the state space representation (\ref{eq:state-space form}) is projected onto the lower dimensional form: 
\begin{equation}
\underbrace{
\left[ \begin{array}{c}
\bar{Y}_{j+1}\\
\delta{\theta}_{j+1} \\
\delta{\bar{Y}}_{j+1} \\
\end{array} 
\right ]}_{\bar{K}_{j+1}}
=
\underbrace{
\left[ \begin{array}{ccc}
I_{l\cdot{P}} & \phi^{+}\widehat{\Gamma^{(P)}K^{(P)}_u}\phi & \phi^{+}\widehat{\Gamma^{(P)}K^{(P)}_y}\phi \\
0_{l\cdot{P}} & 0_{r\cdot{P}} & 0_{l\cdot{P}} \\
0_{l\cdot{P}} & \phi^{+}\widehat{\Gamma^{(P)}K^{(P)}_u}\phi & \phi^{+}\widehat{\Gamma^{(P)}K^{(P)}_y}\phi 
\end{array} 
\right]}_{\bar{A}_j}
\underbrace{
\left[ \begin{array}{c}
\bar{Y}_j \\
\delta{\theta}_j \\
\delta{Y}_j
\end{array} 
\right]}_{\bar{K}_j}
+
\underbrace{
\left[ \begin{array}{c}
\phi^{+}\hat{H}^{(P)}\phi \\
I_{r\cdot{P}} \\
\phi^{+}\hat{H}^{(P)}\phi
\end{array} 
\right]}_{\hat{B}_j}
\delta{\theta}_{j+1}
\, .
\label{eq:state-space form_lower}
\end{equation}
It is worth noting that the size of the projected matrix $\bar{A}\in\mathbb{R}^{12l \times 12l}$ is much smaller than the original matrix $\hat{A}\in\mathbb{R}^{(2l+r)\cdot P \times (2l+r)\cdot P}$. Since in general, $P\gg 2 r$, the order of the optimization problem is significantly reduced by the basis function transformation. 
Furthermore, the transformation guarantees that the input $U_k^{(P)}$ is a smooth signal at the frequencies of interest. 

Based on this, the following quadratic cost function can be minimized to obtain the state feedback gain,
\begin{equation}
J=\sum_{j=0}^{\infty} \left
\Vert
(\bar{K}_{j})^{T}Q\bar{K}_{j}+(\delta\theta_j)^{T}R\delta\theta_j
\right \Vert^2_2
\, ,
\label{eq:cost function}
\end{equation}
where $Q$ and $R$ are user-defined positive-definite weighting matrices for the state and input vectors, respectively, similar to those used for the classical LQR problem. 
Subsequently, the DARE is solved online at each time step, in order to obtain the optimal state feedback gain $K_{f,j}$. Next, the control input vector $\delta\theta_j$ can be formulated based on $K_{f,j}$ according to the state feedback law as,
\begin{equation}
\delta\theta_{j+1}=-K_{f,j}\bar{K}_j
\, .
\label{eq:feadback law}
\end{equation}
Considering that $\delta\theta_{j+1}=\theta_{j+1}-\theta_j$, the projected output update law $\theta_{j+1}$ can be formulated as,
\begin{equation}
\theta_{j+1}=\alpha\theta_j-\beta{K}_{f,j}
\left[ \begin{array}{c}
\bar{Y}_j \\
\delta\theta_j \\
\delta\bar{Y}_j
\end{array} 
\right]
\, ,
\label{eq:update law}
\end{equation}
where $\alpha\in[0,1]$ and $\beta\in[0,1]$ are the tuning parameters to tune the convergence rate of the algorithm. Finally, the input signals $U_k$ can be computed as in {\color{black}equation~\eqref{eq:control input2}}.

Since the Markov matrix $\hat{\Xi}_{k}$ predicted by the RLS optimization in equation~\eqref{eq:Markov parameters3} will take into account the fault dynamics, the adaptive RC law is then formulated according to {\color{black}equations~\eqref{eq:control input2}-\eqref{eq:update law}}, thus leading to load reduction in healthy conditions and to adaptive accommodation for the considered blade and pitch actuator faults. 

%% file: sections/4_simulation.tex
The effectiveness of FTIPC is illustrated through a series of case studies using the 10MW FOWT model in this section. 
Specifically, a comparison between the developed FTIPC, baseline CPC and MBC-based IPC is carried out.
The performance of these control strategies in: 1) blade load reduction, 2) pitch cyclic fatigue loads, 3) effect on FOWT dynamics under different operating conditions and fault scenarios, will be investigated.

\subsection{Model configuration}
The 10MW FOWT model, as introduced in Section \ref{sec:2}, is simulated by the \emph{FAST v8.16} code \cite{Jonkman-2005}. It is connected with \emph{Simulink}, in which the wind turbine torque and pitch controllers are implemented.
In addition, two wind velocities (16m/s, 20m/s), three Turbulence Intensities (TIs, $0\%$, $3.75\%$ and IEC class C) and three fault scenarios (PAD, PAS and blade fault) are simulated, resulting in a total of $3\times 3 \times 2=18$ {\color{black}Load Cases (LCs)}.
Low turbulence is simulated to assess the maximum of the load control potential realized by the approach.
Other chosen wind fields would be the commonly encountered range of operating conditions at the typical FOWT location.
These LCs are displayed in detail in Table \ref{table:load_case}.
The three-dimensional turbulent varying wind field is produced by the TurbSim model\footnote{TurbSim: a stochastic inflow turbulence tool to generate realistic turbulent wind fields. https://nwtc.nrel.gov/TurbSim.\label{TurbSim}}.
The wave conditions under the specific wind speeds are determined by their corresponding conditional probabilistic distribution over North Sea \cite{Li-2015}. 

In each LC, a stretch of 2000\,s is simulated with a fixed discrete time step of $T_s =$0.01\,s, with a predefined fault occurring at 1000\,s.
{\color{black}The parameters $P$ and $p$ are chosen as 100 and 21 respectively.}
%
For the sake of comparisons, three different control strategies, i.e. baseline CPC, MBC-based IPC and FTIPC, are implemented in each LC, thus leading to a total of $3\times18=54$ simulations.


\begin{rem}
{\color{black}
It should be noted that the assumption is made for the case study is that the system parameters vary slowly with time, and the persistency of excitation will lead to the parameters estimates within the bound specified for closed-loop stability and robustness of the system. 
 The bound on parametric uncertainty can be computed according to the analysis in the paper \cite{Martin_1987}.
 In this case, the developed FTIPC will be still closed-loop stable and robust, even though the online system identification does not reach the true values of the system parameters.
 However, when the system parameters vary with time, no guarantee can be made for the stability and robustness of the adaptive repetitive control law. 
 Some approaches have been developed to increase the robustness and stability of the model-based control strategy, such as cautious $\mathcal{H}_2$ optimal control design \cite{Dong_2009}, preview predictive control layer design \cite{Lio_2017}.
 However, it is still an open question for the data-driven FTIPC and will be further investigated in the future.}   
\end{rem}
\subsection{Results and discussions}
The importance of the restricted excitation is discussed firstly by a series of comparison studies. Then the load mitigation and adaptive fault accommodation achieved by FTIPC under various environmental conditions and fault scenarios is illustrated.
\subsubsection{Importance of the restricted excitation}
In order to demonstrate the benefits of the proposed restricted excitation technique, the following comparison is carried out: 
1) FTIPC based on the normal excitation \cite{vanderVeen_2013}, which is marked as $\text{uFTIPC}$ in the following discussions, and
2) FTIPC based on the developed restricted excitation technique.
In the FTIPC case, a restricted exciting signal with a maximum value of $0.1$ is used. 
{\color{black}
On the other hand, the filtered pseudo-random binary signal with a maximum value of $0.25$\,deg is superimposed on top of the pitch angles $U_k^{(P)}$ for system excitation in the $\text{uFTIPC}$ case.}

\begin{figure}[b]
\centering 
\includegraphics[width=1\columnwidth]{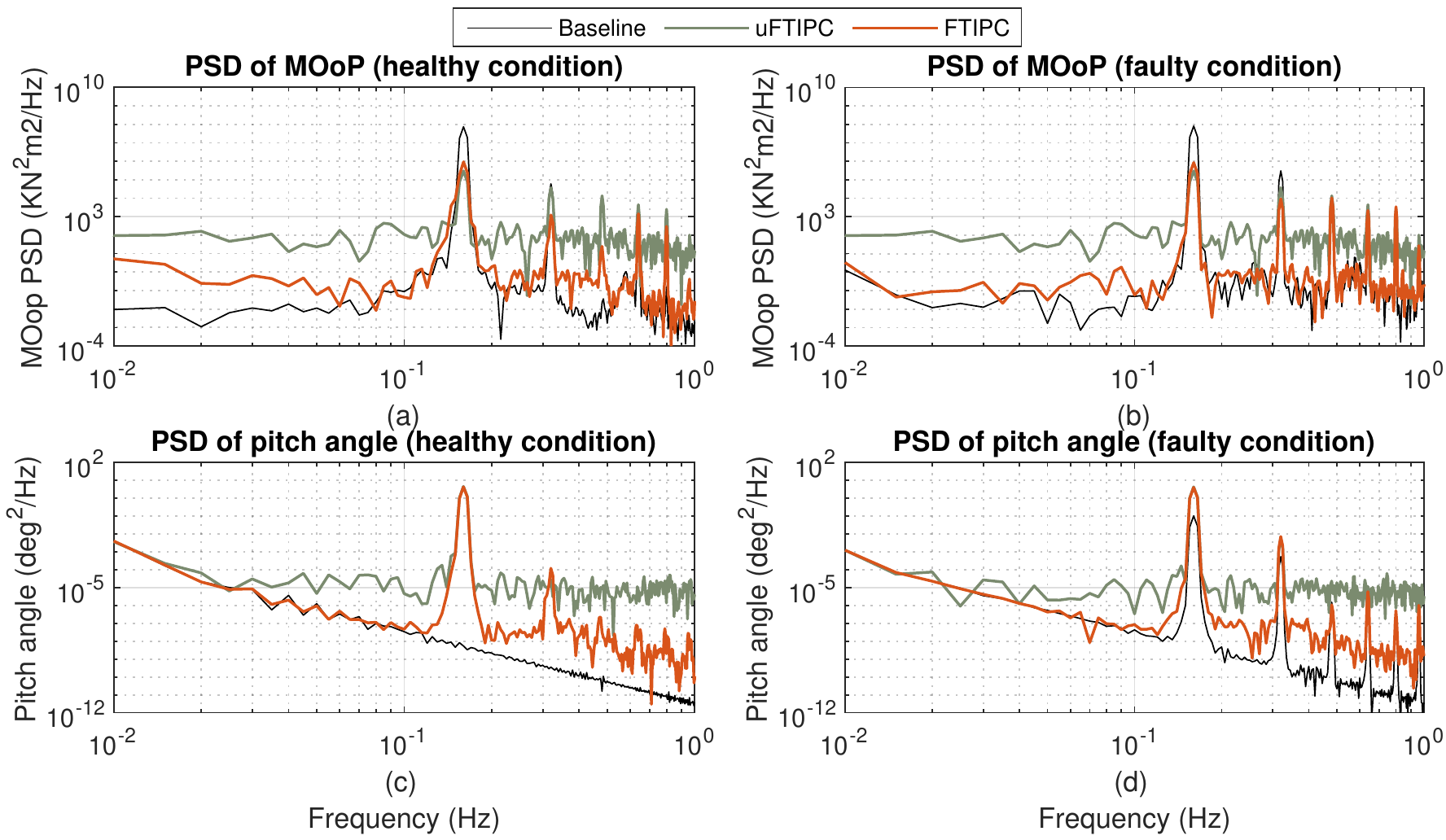}
\caption{PSDs of MOoP and blade 1 pitch angle in LC7 (16m/s wind speed with $0\%$ TI case). Since the pitch angles of blade 2-3 show similar patterns as of blade 1, they are not shown.}
\label{Fig_comp_MOoP} %
\end{figure}

The difference between FTIPC, $\text{uFTIPC}$ and {\color{black}the baseline CPC} is investigated using Power Spectral Density (PSD) functions of MOoP and pitch angles. 
Fig. \ref{Fig_comp_MOoP} presents the results in LC7.
{\color{black}
It should be noted that PSD of MOoPs is computed based on the data from the steady conditions of the case study.
The simulation results derived from 800s-1000s and 1800s-2000s are used to calculate the PSD under healthy and faulty conditions, respectively.}
In general, FTIPC shows a similar load reduction at 1P and 2P frequencies as $\text{uFTIPC}$, however, without much effects on other nominal frequencies (e.g., $10^{-2}$Hz-$10^{-1}$Hz). 
Since the energy of the persistently exciting control input is strictly restricted at 1P and 2P frequencies, the interference on the nominal FOWT dynamics is alleviated. However, $\text{uFTIPC}$ shows higher MOoP and pitch actions in all frequencies. 
For instance, the PSDs of pitch angles derived from $\text{uFTIPC}$ are around {\color{black}$\sim 4.6\cdot 10^{-6}\text{deg}^2/$Hz} at 0.1\,Hz, which is much higher than $\sim 3.5 \cdot 10^{-8}\text{deg}^2/$Hz in the baseline CPC case.
{\color{black}
The reason for this is that all the system dynamics at low frequencies ($\leq$ 1 Hz) are effected by the exciting signal, which is generated by the Butterworth low-pass filter in uFTIPC. This leads to higher PSDs of MOoP and pitch angles.}
Such high pitch actuations in all frequencies may effect the nominal power generation of the operational turbine system and hence increase the failure rate. 

Therefore, it can be concluded that the proposed restricted excitation technique is able to reduce the interference of the exciting signals on the nominal power generation of the wind turbine, by restricting the energy of persistently exciting control input at 1P and 2P frequencies.
Specifically, significant 1P and 2P load reductions are achieved by FTIPC, similar to those derived by $\text{uFTIPC}$, but without much effect on the nominal FOWT dynamics in other frequencies. 
Based on implementation of FTIPC, the load mitigation as well as fault accommodation is realized and discussed in the following subsection. 

\subsubsection{Load mitigation and fault accommodation}
\begin{figure}[b]
\centering 
\includegraphics[width=1\columnwidth]{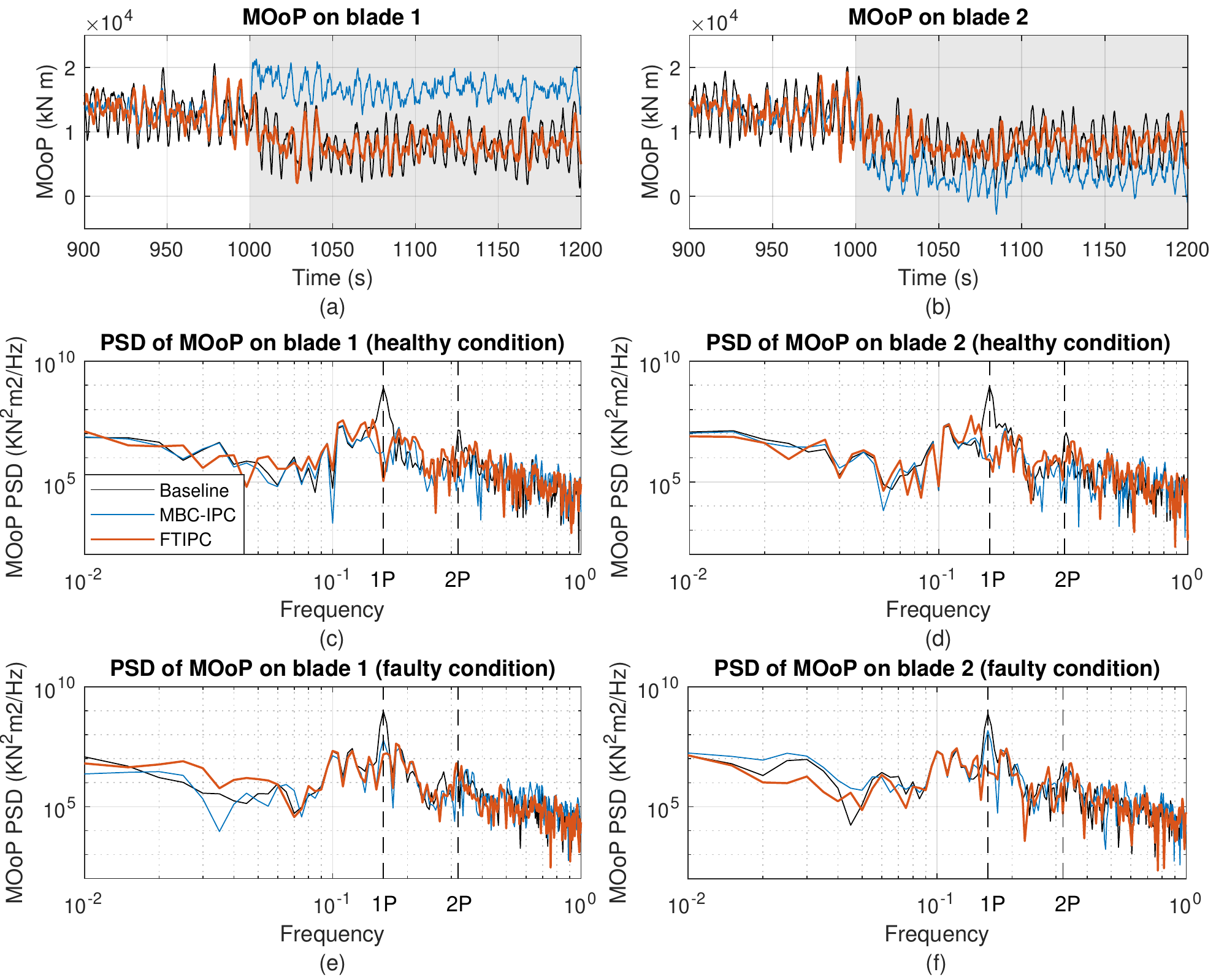}
\caption{MOoP on the blade root in LC2 (16m/s wind speed with $3.75\%$ TI case) with a PAD occurring at 1000\,s. Time periods of fault scenarios are indicated by a grey background. Blade 3 is not shown since {\color{black}it is the faulty blade}.}
\label{Fig_MOoP_375_PAD} %
\end{figure}

The effectiveness of the proposed FTIPC in load mitigation and fault accommodation is illustrated through a series of comparison studies.
Figure~\ref{Fig_MOoP_375_PAD} presents the MOoPs on the blade root in LC2 with a PAD fault occurring at 1000s.
Under the healthy condition before 1000\,s, both MBC-based IPC and FTIPC show effective load reduction at 1P and 2P frequencies. 
%
Since the Markov parameters of SPRC are identified online for each blade, the proposed FTIPC would be adapted to the faulty conditions, taking into account the PAD fault after 1000\,s.
With the inclusion of the fault dynamics in the linear model, the adaptive RC law is formulated to accommodate the PAD-induced blade loads. As visible in Figure~\ref{Fig_MOoP_375_PAD}(a-b), a significant load reduction has been achieved by FTIPC under faulty conditions, which prevents the PAD fault from deteriorating.
The variations of MOoP are reduced by more than ${\color{black}\sim 55\,\%}$. This leads to an effective fault accommodation of the PAD fault. 

\begin{figure}[t]
\centering 
\includegraphics[width=1\columnwidth]{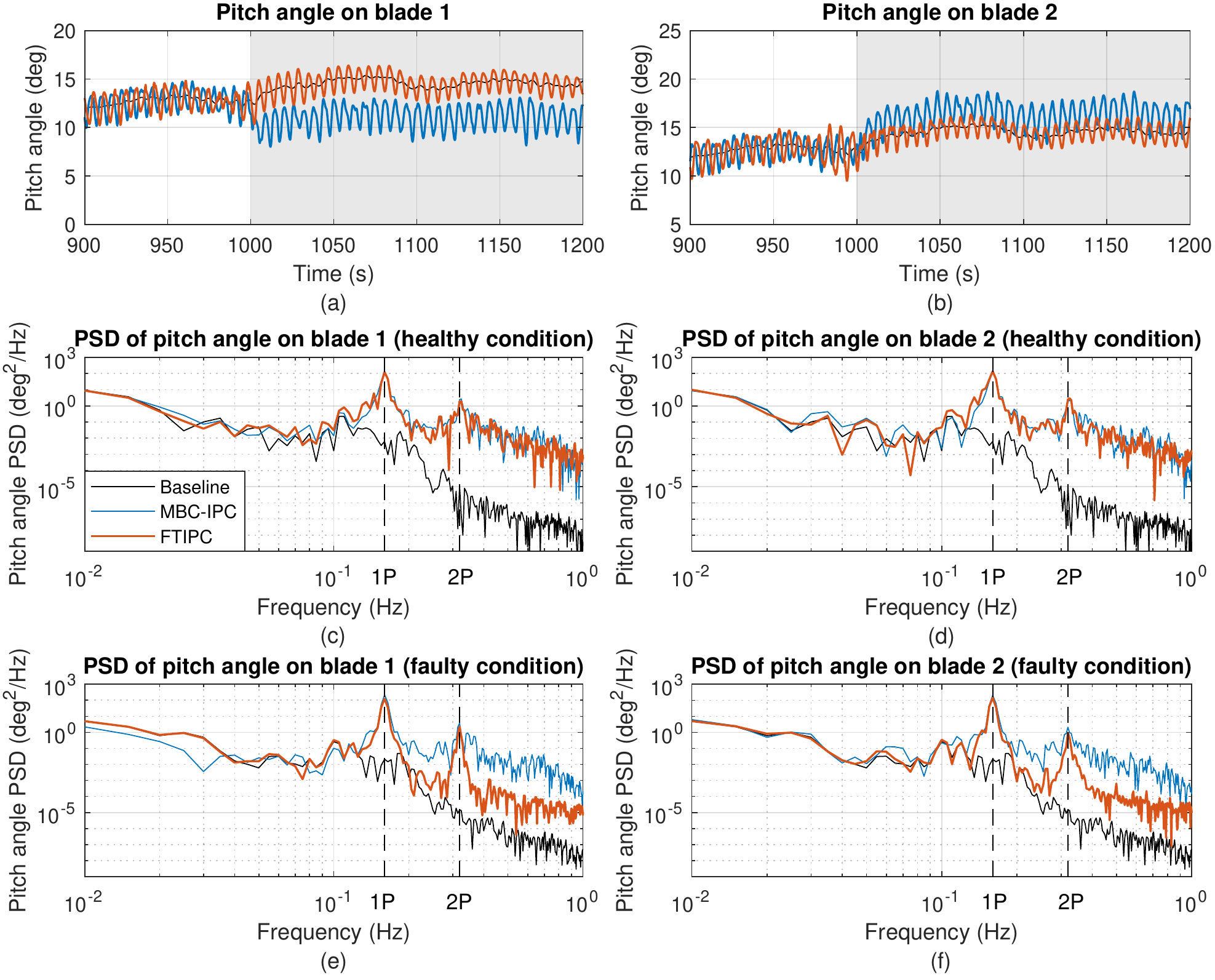}
\caption{Pitch angle in LC2 (16m/s wind speed with $3.75\%$ TI case) with a PAD occurring at 1000\,s. Time periods of fault scenarios are indicated by a grey background in (a-b). Time periods of fault scenarios are indicated by a grey background. Blade 3 is not shown since {\color{black}it is the faulty blade}.}
\label{Fig_pitchangle_375_PAD} %
\end{figure}

MBC-based IPC, on the other hand, is unable to address the PAD fault.
It imposes pitch inputs with a mean value different from the one of the baseline CPC and SPRC-based IPC, as illustrated in Figure~\ref{Fig_pitchangle_375_PAD}. This makes the mean of MOoP deviate from baseline CPC and FTIPC case, which further deteriorates the aerodynamic asymmetry.
{\color{black}This is caused by the contamination of the signals from the Coleman transformation utilized in MBC-based IPC due to the faulty pitch actuator.}
{\color{black}The performance of FTIPC and the baseline controller} can also be discerned in PSDs in Figure~\ref{Fig_MOoP_375_PAD}(e-f).
{\color{black}As visible}, FTIPC {\color{black}still} reduces the maximum PSD of MOoP into $\sim 10^6$\,kN$^2$ m$^2$/Hz {\color{black}at 1P frequency under faulty conditions}.
Therefore, FTIPC shows the best performance in the PAD fault scenario.

\begin{figure}[t]
\centering 
\includegraphics[width=1\columnwidth]{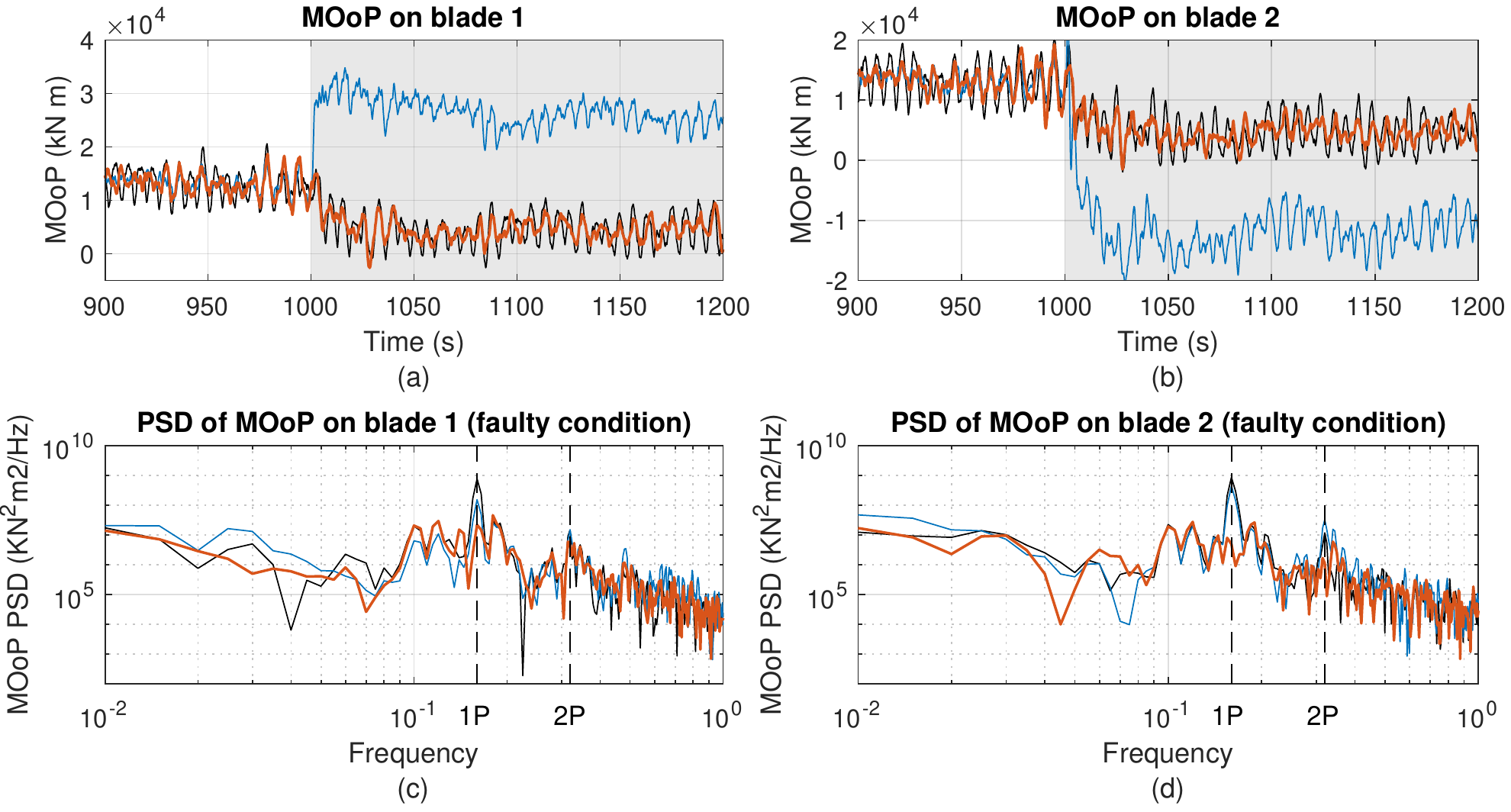}
\caption{MOoP on the blade root in LC5 (16m/s wind speed with $3.75\%$ TI case) with a PAS occurring at 1000\,s. Time periods of fault scenarios are indicated by a grey background. Blade 3 is not shown since {\color{black}it is the faulty blade}.}
\label{Fig_MOoP_375_PAS} %
\end{figure}

\begin{figure}[t]
\centering 
\includegraphics[width=1\columnwidth]{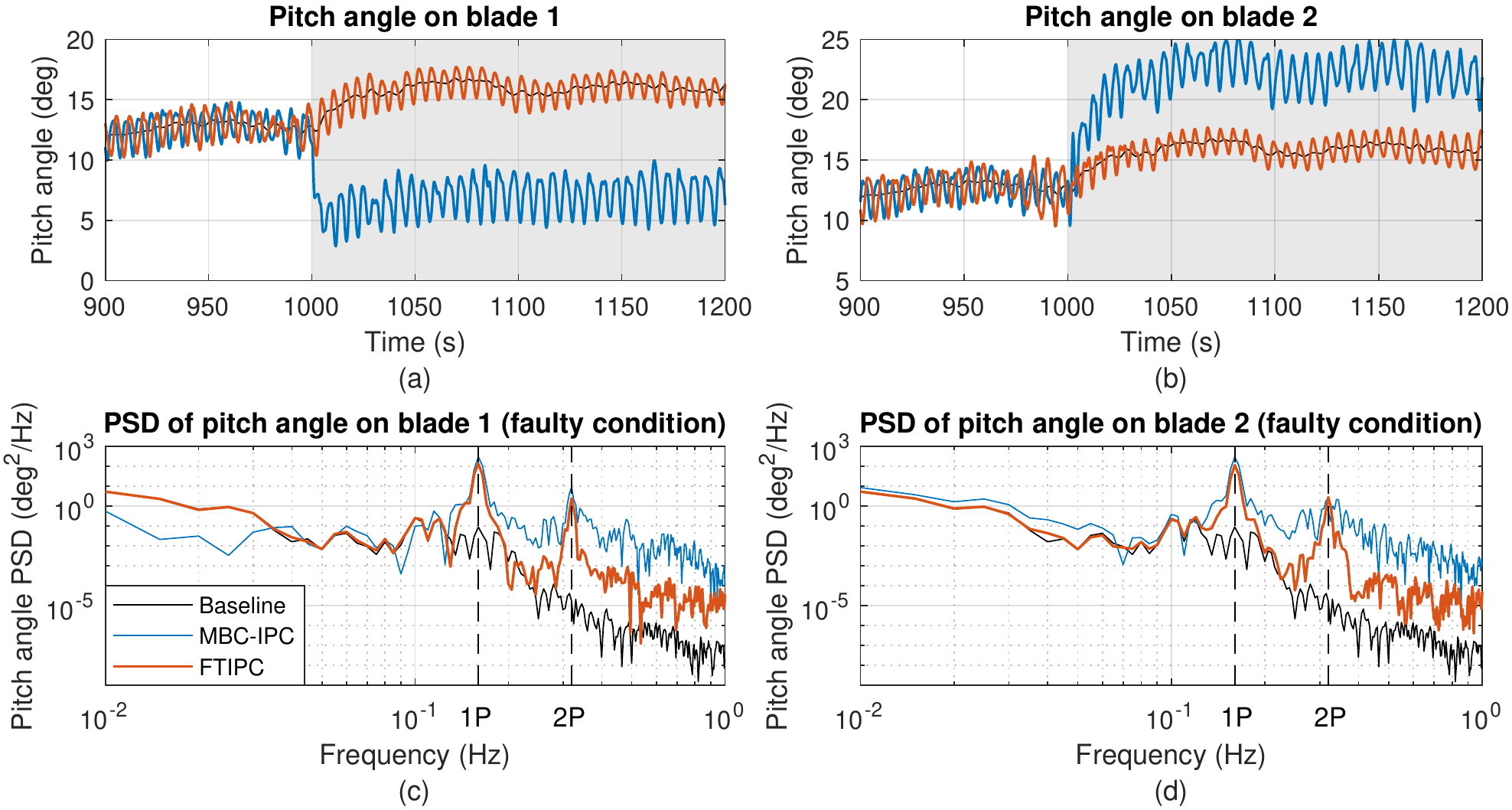}
\caption{Pitch angle in L5 (16m/s wind speed with $3.75\%$ TI case) with a PAS occurring at 1000\,s. Time periods of fault scenarios are indicated by a grey background in (a-b). Time periods of fault scenarios are indicated by a grey background. Blade 3 is not shown since {\color{black}it is the faulty blade}.}
\label{Fig_pitchangle_375_PAS} %
\end{figure}

In case of the PAS fault, the MOoP on blade $3$ is unable to be minimized since its pitch actuator is stuck after 1000\,s.
The proposed FTIPC outperforms in load mitigation in this fault scenario, which is similar as in PAD case. 
Some results are depicted in Figures~\ref{Fig_MOoP_375_PAS} and \ref{Fig_pitchangle_375_PAS}.
It can be discerned that the proposed FTIPC reduces PAS-induced blade loads and alleviate the aerodynamic asymmetry effectively.
For instance, the variations of MOoP are reduced by more than ${\color{black}\sim 52\,\%}$ under faulty conditions.
However, MBC-based IPC is infeasible to deal with the PAS fault.
Evident deviations of the mean pitch inputs between MBC-based IPC and baseline CPC are observed after 1000\,s. 
Such erroneous pitch signals lead to the increased MOoP, and make the aerodynamic asymmetry even worse than in PAD case.
%
For pitch actuator faults, it can be concluded that, significant load mitigation, especially an effective fault accommodation, can be achieved by the proposed FTIPC. 

\begin{figure}
\centering 
\includegraphics[width=1\columnwidth]{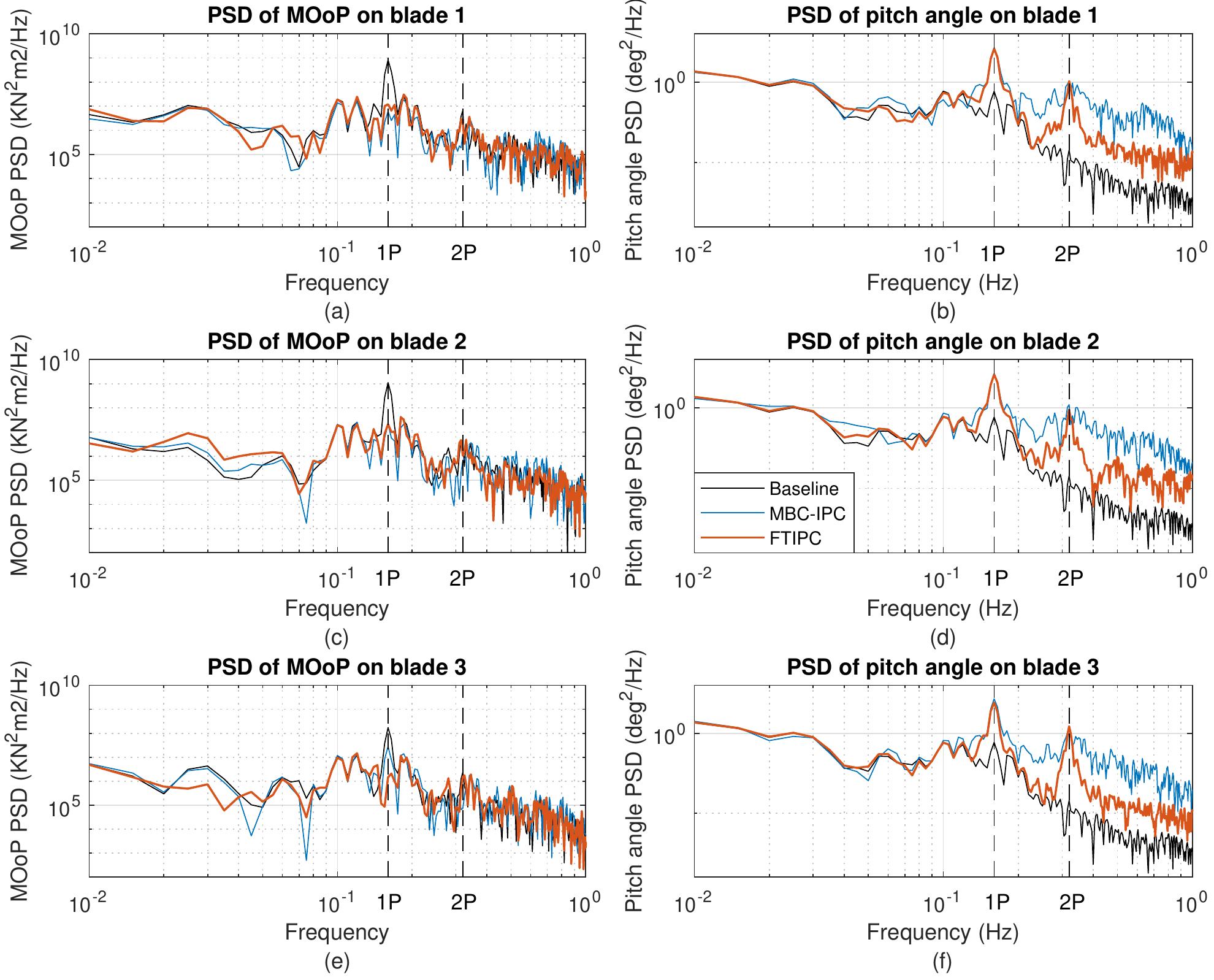}
\caption{PSDs of MOoP and pitch angles in LC8 (16m/s wind speed with $3.75\%$ TI case) with a blade fault occurring at 1000\,s. }
\label{Fig_375_BF} %
\end{figure}

Finally,  the fault accommodation of the blade fault is also taken into consideration by FTIPC.
{\color{black}Figure~\ref{Fig_375_BF} depicts PSDs of MOoP and pitch inputs in LC8 where blade 3 suffers from faults at 1000\,s.}
MBC-based IPC is not an effective way to reduce MOoP on each blade, hence resulting in higher spectral values at 1P and 2P frequencies.
On the other hand, the proposed FTIPC shows the best performance in fault accommodation. 
Since the blade fault dynamics are successfully learned by SPRC, the adaptive fault-tolerant RC law is formulated in Figure~\ref{Fig_375_BF}(b,d,f) to accommodate the fault. 
As a result, MOoP {\color{black}at} each blade {\color{black}root}, which is in essence induced by the blade fault, is reduced, as shown in Figure~\ref{Fig_375_BF}(a,c,e).
In conclusion, the proposed FTIPC is an effective approach to mitigate MOoP on both healthy and faulty blades in case of the blade fault.

Similar results are found in other considered LCs. For the purpose of comparisons, all these results are summarized in Table~\ref{table:load_reduction}.
In detail, the {\color{black}indicator, relative Standard Deviation (rSD) of MOoP}, is used to quantify the performance of the analyzed three control strategies. 
{\color{black}rSD denotes the SD reduction of MOoP achieved by the IPC strategies compared to the baseline controller, which is calculated as,}
\begin{equation}
{\color{black}
    \text{rSD}_{\text{IPC}} = (\text{SD}_{\text{baseline}}-\text{SD}_{\text{IPC}})/\text{SD}_{\text{baseline}}\,,
    }
  \end{equation}
{\color{black}
where $\text{SD}_{\text{baseline}}$ and $\text{SD}_{\text{IPC}}$ denote SD of MOoP from baseline controller and IPC strategies, respectively.
From Table~\ref{table:load_reduction}}, the proposed FTIPC {\color{black}in general} shows the best performance in fault accommodation for all considered LCs. 
The most evident load mitigation appears in blade fault cases, most of which showing $\sim 60\,\%$.
One plausible explanation is that all actuators are still working in case of blade faults, which make FTIPC be the most effective way to accommodate blade faults.
{\color{black}During PAS fault cases, more than $52\,\%$ MOoP is reduced by FTIPC in LC4-LC6 and LC13-LC15. 
This actually implies a significant load reduction in such a severe fault scenario. }
Averaging over all the cases, the proposed method leads to an effective fault-induced load mitigation by ${\color{black}58.44\,\%}$.
MBC-based IPC, on the other hand, fails to formulate proper pitch inputs to accommodate blade and pitch actuator faults due to the contaminated Coleman transformation. 
Much difference of the MOoP standard deviation between blade 1 and 2 is observed in this control strategy.
For example, blade 1 shows $42.42\%$ MOoP reduction while blade 2 is only $8.86\%$ MOoP reduction in LC4-LC6, which results in an increased aerodynamic asymmetry.
Moreover, the values in italic in Table~\ref{table:load_reduction} imply that the loads are in fact increased by MBC-based IPC compared to the baseline CPC. 
Therefore, in comparison, the conventional MBC-based IPC fails to deal with blade and pitch actuator faults and makes the faulty condition even worse, while the proposed FTIPC is a feasible way to accommodate all considered faults in different LCs.

\subsubsection{Assessment of the cyclic fatigue loads on the actuators}
\begin{figure}
\centering 
\includegraphics[width=1\columnwidth]{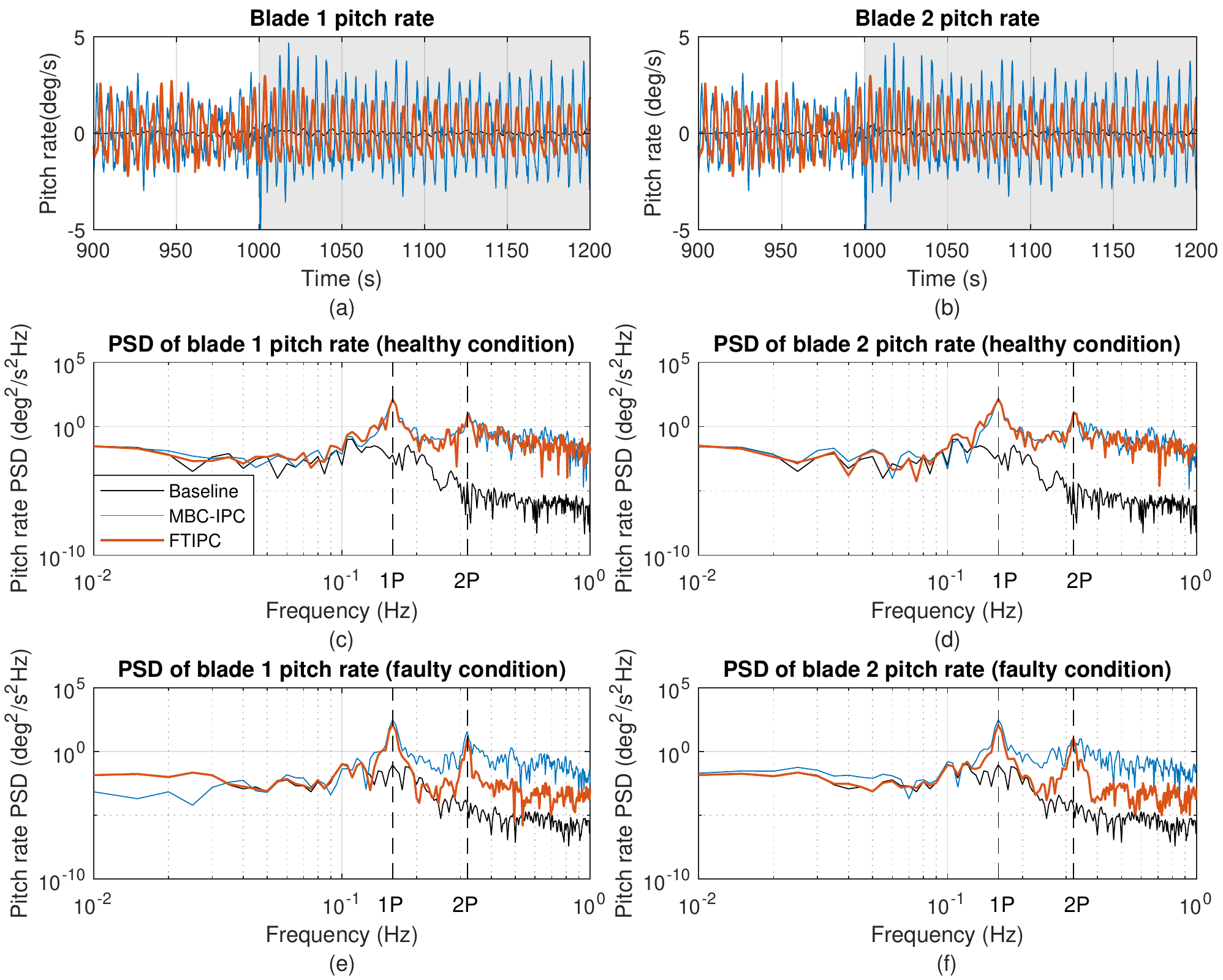}
\caption{Pitch rate in LC5 (16m/s wind speed with $3.75\%$ TI case) with a PAS occurring at 1000\,s. Time periods of fault scenarios are indicated by a grey background. Blade 3 is not shown since {\color{black}it is the faulty blade}.}
\label{Fig_pitch_rate_375_PAS_PSD} %
\end{figure}

In addition to the fault accommodation, another benefit that can be anticipated is that the proposed FTIPC requires relatively low pitch activities compared to MBC-based IPC. 
Figure~\ref{Fig_pitch_rate_375_PAS_PSD} presents the pitch action of blade 1-2 in L5. Since the control input is restricted at 1P and 2P frequencies, FTIPC shows much slower speed of pitch angle than MBC-based IPC. 
The use of FTIPC reduces pitching activities in this case by ${\color{black}38.42\%}$.
However, MBC-based IPC covers a much broader band of frequencies as shown in Figure~\ref{Fig_pitch_rate_375_PAS_PSD}({\color{black}e-f}), which actually increases pitch activities.
Such increased pitch actions may cause the rise of the risk of cyclic fatigue failure on the actuators, thus introduces inferences on nominal dynamics of the wind turbine. 
Thanks to the reduced pitch activities in FTIPC case, the cyclic fatigue loads are alleviated substantially.

The cyclic fatigue loads on the actuators can typically be quantified by the Actuator Duty Cycle (ADC) \cite{Bottasso_2013}, which is a widely-used criterion to estimate the lifespan of pitch actuators. In this paper, ADC for all LCs are given in Table~\ref{table:ADC}.
Compared to MBC-based IPC, the proposed FTIPC shows much lower ADC of both blade 1 and 2.
Averaging over all the cases, the ADC is reduced by FTIPC by ${\color{black}25.11\,\%}$ {\color{black}compared with MBC-IPC}.
The most evident reduction of ADC, showing a maximum of ${\color{black}38.19\,\%}$, appears in {\color{black}LC13-LC15}. 
Thus, the benefit can be seen that only ${\color{black}\sim 12\,\%}$ ADC is requested by FTIPC, however, substantial load mitigation and fault accommodation are achieved.
This implies that FTIPC is able to achieve significant load mitigation and fault accommodation with lower pitch activities.
Since the ADC, indicating the cyclic fatigue loads on the actuators, is substantially reduced, this approach will help to increase the reliability of the pitch system.

\subsubsection{Effect on power generation and platform motions}
\begin{figure}
\centering 
\includegraphics[width=1\columnwidth]{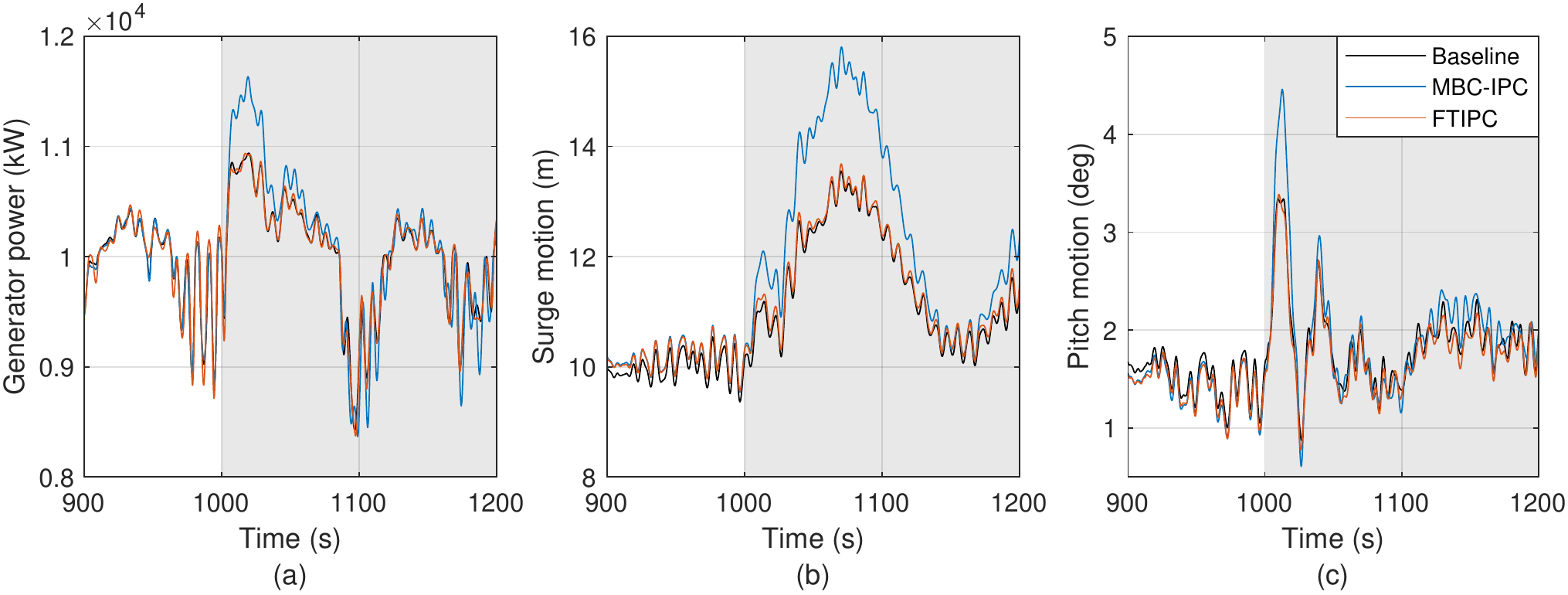}
\caption{Generator power of the wind turbine, surge and pitch motions of the floating platform in LC5 (16m/s wind speed with $3.75\%$ TI case) with a PAS occurring at 1000\,s. (a) Generator power, (b) Surge motion, (c) Pitch motion. Time periods of fault scenarios are indicated by a grey background.}
\label{Fig_PW_motion375} %
\end{figure}

Finally, the global FOWT dynamics under different control strategies are of concern and discussed in this section. 
Figure~\ref{Fig_PW_motion375} shows the {\color{black}generator power} and platform motions (i.e. surge, pitch) in LC5.
It can be seen that excessive generator power and a big increase of platform motions are observed at fault time, i.e. 1000\,s.
The worst response appears in the MBC-based IPC case, showing a generator power of more than 11.5\,MW, surge of more than 15.9\,m and pitch of more than 4.5\,deg.
{\color{black}Such increased power fluctuation and platform motions will result in reduced reliability and stability of the turbine system.}
On the other hand, both baseline CPC and FTIPC show much lower responses to the faults.
In the baseline CPC, each healthy blade keeps the same pitch angle due to the single pitch actuator, which potentially avoid the effect of the aerodynamic asymmetries on the rotor.
Regarding FTIPC, this approach is able to rapidly alleviate PAS-induced aerodynamic asymmetries on the rotor blades, thus minimizing the {\color{black}power variations} and platform motions down to the baseline CPC level. 
Other LCs show similar results and are therefore omitted for brevity. 

Based on the results presented in this section, it can be concluded that
the proposed FTIPC is able to alleviate the aerodynamic asymmetries on the rotor blades down to the level comparable with the baseline CPC.
{\color{black}As a consequence, low power variations and platform motions can be guaranteed when the proposed load mitigation strategy is deployed. 
Therefore, this approach} makes it possible to keep basic FOWT dynamics and continue power generation before necessary maintenance is carried out in case of the blade and pitch actuator faults.

%% file: sections/5_conclusion.tex
Blade and pitch actuator faults pose a challenge to the control engineer as the conventional Individual Pitch Control (IPC) does not work in these faulty cases.
In this paper, a Fault-Tolerant Individual Pitch Control (FTIPC) is developed based on the Subspace Predictive Repetitive Control (SPRC) approach to address the situation of blade and pitch actuator faults in Floating Offshore Wind Turbines (FOWTs). 
This control strategy integrates online subspace identification with repetitive control to minimize the blade loads under operational conditions. 
By identifying the Markov matrix for each blade online, FTIPC is able to accommodate faults in an adaptive way.
Furthermore, this paper introduces a novel technique where the energy of the persistently exciting control input is limited to the specific frequencies. This technique reduces the interference of the exciting signals on the nominal dynamics of wind turbines. 
The effectiveness of the developed FTIPC is demonstrated on a 10MW FOWT model through a series of case studies, where different operating conditions and fault scenarios are considered. 

These simulation cases show that
the conventional Multi-Blade Coordinate (MBC)-based IPC fails in case of blade and pitch actuator faults.
The proposed FTIPC is an effective way to attain load mitigation and adaptive fault accommodation, via a restricted persistently exciting control input.
A comparison with the baseline Collective Pitch Control (CPC) and MBC-based IPC indicates that FTIPC outperforms in all considered fault scenarios and operating conditions.
This approach shows the best performance in fault accommodation and pitch Actuator Duty Cycle (ADC) among the considered control strategies.
Averaging over all the cases, FTIPC achieves a reduction of the fault-induced loads by ${\color{black}58.44\%}$. Moreover, the use of FTIPC reduces ADC by ${\color{black}25.11\%}$ compared to MBC-based IPC.

Therefore, it can be concluded that the proposed scheme is a significant improvement with respect to existing control strategies, since it can be adapted to faulty conditions and continue to guarantee load control performance in case of blade and pitch actuator faults.
As a result, the aerodynamic asymmetries on the rotor caused by the faults, are alleviated down to the baseline CPC level.
{\color{black}Considering that significant load mitigation is attained by FTIPC in both severe and non-severe fault scenarios, it will help} to keep the basic performance of the faulty FOWT, and hence makes it possible to continue power production before necessary maintenance is performed.

Based on the case studies in this paper, the advantages can be seen that the developed FTIPC, shows high adaptability and low pitch ADC for load reduction under healthy and faulty conditions. It is able to consider both severe and non-severe faults by learning the faulty system dynamics automatically.
{\color{black}Since it is a fully data-driven adaptive approach, it is also possible to handle the coexistence of multiple faults and other types of faults.
Future work will include more realistic and complex fault scenarios.
Moreover, the model uncertainty may exert negative effects on control performance.
In this respect, the stability of the SPRC algorithm and the safety constraints of the pitch actuators should be taken into account.
All of them will be further investigated in the future.}



%% file: sections/Tables.tex
\newpage
\begin{table}[h]
\centering
{
\caption{Specification of the 10MW FOWT model.\label{table:FOWT}}

\begin{tabular}{ll}
\hline 
\textbf{Parameter}     & \textbf{Value}         \\  \hline 
\textbf{Turbine system}           \\
Rating           & 10\,MW          \\
Rotor orientation, configuration  & Upwind, 3 blades            \\
Pitch control         & Variable speed, baseline and IPC       \\
Drivetrain      & Medium speed, multiple stage gearbox          \\
Rotor, hub diameter & 178.3\,m, 5.6\,m \\
Hub height & 119\,m \\
Cut-in, rated, cut-out wind speed & 4\,m/s, 11.4\,m/s, 25\,m/s \\
Cut-in, rated rotor speed & 6\,rpm, 9.6\,rpm \\
Rated tip speed & 90\,m/s \\ \hline
\textbf{Floating platform}  \\ 
Total height and draft & 66\,m, 56\,m \\
Distance from the tower center-line & 26\,m \\
Single column diameter & 15\,m \\
Column elevation above Sea Water Level (SWL) & 10\,m \\
Elevation of tower base above SWL & 25\,m \\
Water displacement & 29497.7 \,m$^3$ \\ \hline
\textbf{Mooring lines} \\ 
Number of lines & 3 \\
Line angles from upwind direction & 0\,deg, 120\,deg, 240\,deg \\
Anchor depth and radius & 180\,m, 599.98\,m \\
Fairleads above SWL & 8.7\,m \\
Fairleads radius & 47.181\,m \\
Line diameter & 0.18\,m \\
Total length & 707\,m \\
Mass/length in air & 594\,kg/m \\ 
\hline  
\end{tabular}}
\end{table}

\begin{table}[h]
\centering
{
\caption{Fault scenarios and environmental conditions in all LCs.\label{table:load_case}}
\begin{tabular}{llll}
\hline
\textbf{Load case} & \textbf{Wind speed $U_m$} & \textbf{Turbulence intensity}   & \textbf{Fault scenarios}\\ \hline

\multirow{3}*{$1-3$}  & 
\multirow{3}*{16\,m/s}      & 
$0\,\%$ TI  & 
\multirow{3}*{$\vartheta^{x}_{\text{PAD}}=0.5$}\\ 
 & & $3.75\,\%$ TI & 
\\ 
 & & IEC class C &

\\\hline
\multirow{3}*{$4-6$}  &
\multirow{3}*{16\,m/s} & 
$0\,\%$ TI
&
\multirow{3}*{$\vartheta^{x}_{\text{PAS}}=0\,\text{deg}$}\\
 & & $3.75\,\%$ TI & 
 \\
 & & IEC class C &

\\ \hline

\multirow{3}*{$7-9$}  &
\multirow{3}*{16\,m/s} & 
$0\,\%$ TI
&
\multirow{3}*{$\vartheta^{x}_{\text{B}}=0.2\cdot \vartheta^{x}$}\\
 & & $3.75\,\%$ TI & 
 \\
 & & IEC class C &

\\ \hline

\multirow{3}*{$10-12$}  & 
\multirow{3}*{20\,m/s}      & 
$0\,\%$ TI  & 
\multirow{3}*{$\vartheta^{x}_{\text{PAD}}=0.5$}\\ 
 & & $3.75\,\%$ TI & 
\\ 
 & & IEC class C &

\\\hline
\multirow{3}*{$13-15$}  &
\multirow{3}*{20\,m/s} & 
$0\,\%$ TI
&
\multirow{3}*{$\vartheta^{x}_{\text{PAS}}=10\,\text{deg}$}\\
 & & $3.75\,\%$ TI & 
 \\
 & & IEC class C &

\\ \hline

\multirow{3}*{$16-18$}  &
\multirow{3}*{20\,m/s} & 
$0\,\%$ TI
&
\multirow{3}*{$\vartheta^{x}_{\text{B}}=0.2\cdot \vartheta^{x}$}\\
 & & $3.75\,\%$ TI & 
 \\
 & & IEC class C &

\\ \hline
\end{tabular}}

\footnotesize
\vspace{0.1cm}
\color{black}{
*In this study, it is assumed that only the third blade and its pitch actuator are subjected to the faults in all considered cases.}
\end{table}

\begin{table}
\setlength{\tabcolsep}{3.5mm}
{
\caption{Comparisons of MOoP reductions with different control strategies under faulty conditions*. \label{table:load_reduction}}
\begin{tabular}{lcccccc}
\hline 
\textbf{MOoP}
& $\text{LC}1-\text{LC}3$   & $\text{LC}4-\text{LC}6$   & $\text{LC}7-\text{LC}9$  &$\text{LC}10-\text{LC}12$ & $\text{LC}13-\text{LC}15$  & $\text{LC}16-\text{LC}18$   
\\ \hline
\textbf{Blade} $1$ \\ 
MBC-IPC (\%)  & 47.85            & 42.42  & 64.38    &  55.56 & 34.98        & 59.51
\\
FTIPC (\%)    & {\color{black}59.65}  & {\color{black}55.11}   & {\color{black}60.05}   &  {\color{black}54.34} & {\color{black}52.42}        & {\color{black}53.24}
\\ 

\textbf{Blade} $2$ \\ 
MBC-IPC (\%) & \textit{-6.35}   & 8.86   & 53.94    &  32.01 & \textit{-1.31} & 57.32
\\
FTIPC (\%)   & {\color{black}56.74}  & {\color{black}58.63}  & {\color{black}63.09}   &  {\color{black}53.70} & {\color{black}54.21}   & {\color{black}59.10}
\\ 
\textbf{Rotor hub} \\ 
MBC-IPC (\%) & 19.31  & 44.82   &  70.95 &  
62.90 & 38.29   & 73.18
\\
FTIPC (\%)    & {\color{black}48.41}   & {\color{black}57.56}   &  {\color{black}70.27} &
{\color{black}66.77} & {\color{black}58.64}  & {\color{black}69.92}
\\ \hline 
\end{tabular}}

\footnotesize
\vspace{0.1cm}
*These results are calculated based on the data $1800\,\text{s}- 2000\,\text{s}$. They represent the percentage of changes in MOoP with respect to the baseline CPC case. The negative values indicate that the load increases and the controller has worse performance than the baseline one.
\end{table}

\begin{table}
\setlength{\tabcolsep}{3mm}
{
\caption{ADC of the blade pitch actuators for the FOWT in all LCs*.\label{table:ADC}}
\begin{tabular}{lcccccc}
\hline
\textbf{ADC} 
& $\text{LC}1-\text{LC}3$   & $\text{LC}4-\text{LC}6$   & $\text{LC}7-\text{LC}9$  &$\text{LC}10-\text{LC}12$ & $\text{LC}13-\text{LC}15$  & $\text{LC}16-\text{LC}18$   
\\ \hline
\textbf{Blade} $1$ \\ 

Baseline CPC ($\%$)  
& 0.67  & 0.63  & 0.74  & 0.99  & 1.00  & 1.06  %
\\ 
MBC-IPC ($\%$)
& 14.47  & 15.82  & 11.48  & 15.55 & 18.78  & 13.98  %
\\ 
FTIPC ($\%$)   
& {\color{black}11.03}  & {\color{black}10.90}  & {\color{black}9.60}  & {\color{black}12.30} & {\color{black}12.36}  & {\color{black}11.29}  %
\\ %

\textbf{Blade} $2$ \\ 
Baseline CPC ($\%$)     
& 0.67  & 0.63  & 0.74  & 0.99 & 1.00  & 1.06  %
\\ 
MBC-IPC ($\%$) 
& 14.57  & 17.35  & 12.07  & 16.56 & 20.22  & 13.69 %
\\ 
FTIPC ($\%$)   
& {\color{black}11.01}  & {\color{black}10.16}  & {\color{black}10.83}  & {\color{black}12.50} & {\color{black}11.69}  & {\color{black}11.93}  %
\\ %
\hline 

\end{tabular}}

\footnotesize
\vspace{0.1cm}
*The faulty conditions cover the data from $1800\,\text{s}- 2000\,\text{s}$ for the ADC calculations.
\end{table}